\documentclass[11pt,a4paper]{article}
 
\usepackage[pdftex]{graphicx}
\usepackage[cmex10]{amsmath}
\usepackage{amsfonts}
\usepackage{amssymb}
\usepackage{amsthm}
\usepackage{siunitx}

\usepackage{algorithm}
\usepackage{algpseudocode}
\usepackage[hidelinks=true]{hyperref}

\usepackage{xfrac}
\usepackage{enumerate}

\usepackage{subfigure}
\usepackage[font=footnotesize ]{caption}
\renewcommand{\figurename}{\bfseries{Fig.}}
\renewcommand{\tablename}{\bfseries{Table}}

\usepackage{multirow}
\usepackage{rotating}
\usepackage{makecell}
\usepackage{booktabs}
\usepackage{tabularx,ragged2e}
\usepackage{longtable}
\usepackage{anysize}
\marginsize{2.5cm}{2.5cm}{2.5cm}{2.5cm}

\usepackage{xcolor}
\definecolor{correct}{rgb}{0.77, 0.12, 0.23} 
\definecolor{revise}{rgb}{0.0, 0.34, 0.25} 
\definecolor{insert}{rgb}{0.0, 0.2, 0.4} 

\usepackage{setspace}
\usepackage{parskip}
\usepackage[pagewise]{lineno}

\usepackage{lipsum}

\usepackage{authblk}

\title{\textbf{Design and Characterization of Effective Solar Cells}}
\author[1]{Varun~Ojha\thanks{{\footnotesize VO and GJ contributed equally to this work.\\
		Cite this work as:~\\Ojha, V., Jansen, G., Patan\`{e}, A. et al. Design and characterization of effective solar cells. \textit{Energy Systems} (2021)}}}
\author[2]{Giorgio Jansen}	
\author[3]{Andrea Patan\`{e}}
\author[4]{Antonino La Magna}
\author[5]{Vittorio Romano}
\author[2,6]{Giuseppe Nicosia}

\affil[1]{Department of Computer Science,  University of Reading, Reading, UK}
\affil[2]{Systems Biology Centre, University of Cambridge, Cambridge, UK}
\affil[3]{Department of Computer Science, University of Oxford, Oxford UK}
\affil[4]{National Research Council-Institute for Microelectronics and Microsystems, Catania, Italy}
\affil[5]{Department of Mathematics \& Computer Science, University of Catania, Italy}
\affil[6]{Department of Biomedical \& Biotechnological Sciences, University of Catania, Catania, Italy}

\date{}
\begin{document}

	\maketitle              
	\begin{abstract}
		We propose a two-stage multi-objective optimization framework for full scheme solar cell \textit{structure design and characterization}, \textit{cost minimization} and \textit{quantum efficiency maximization}. 
		We evaluated structures of 15 different cell designs simulated by varying material types and photodiode doping strategies.
		At first, non-dominated sorting genetic algorithm~II (NSGA-II) produced Pareto-optimal-solutions sets for respective cell designs. 
		Then, on investigating quantum efficiencies of all cell designs produced by NSGA-II, we applied a new \textit{multi-objective optimization algorithm~II} (OptIA-II) to discover the Pareto fronts of select (three) best cell designs. 
		Our designed OptIA-II algorithm improved the quantum efficiencies of all select cell designs and reduced their fabrication costs. 
		We observed that the cell design comprising an optimally doped zinc-oxide-based transparent conductive oxide (TCO) layer and rough silver back reflector (BR) offered a quantum efficiency ($Q_e$) of $0.6031.$
		%
		Overall, this paper provides a \textit{full characterization} of cell structure designs. 
		It derives relationship between quantum efficiency, $Q_e$ of a cell with its TCO layer's doping methods and TCO and BR layer's material types. 
		Our solar cells design characterization enables us to perform a cost-benefit analysis of solar cells usage in real-world applications.
		
	\end{abstract}

	\section{Introduction 
		\label{sec:_intro_sloarCell}}
	Sustainable energy demand of 21st century comes from green energy production methods like harvesting energy from nature: solar, water, and wind.
	Of these, solar light harvesting is pervasive and is the most agile in terms of installation. 
	
	Solar cells are typically categorized as photovoltaic~\cite{hagfeldt2000molecular}, thermophotovoltaic~\cite{bermel2010design}, 
	or nanophotonic thermophotovoltaic~\cite{lenert2014nanophotonic} type cells. 
	Since solar energy is the most used green energy method, many research works (e.g.,~\cite{shah2004thin,atwater2011plasmonics,geisz200840}) have been done on solar cell design and cell structure optimization to improve cells light-harvesting efficiency and solar energy production capacity maximization.

	In our previous work~\cite{patane2018enhancing}, we performed simulation and experiments to improve thin-film photovoltaic type solar cell design and its light-harvesting efficiency. 
	In this paper, we aim to improve upon our previous results~\cite{patane2018enhancing} and extend our framework to investigate a \textit{full characterization of optimal solar cell design}. 
	Thus, we performed full scheme solar cell design simulations and investigated their Pareto surfaces.
	We evaluated various solar cell compositions and material combinations for designing different solar cell structures.  

    We formulated ``solar cell structure design problem'' and its optical simulations for cells quantum efficiency improvement as a \textit{multi-objective optimization} (MOO) problem~\cite{Nicosia06,Nicosia08}.
    We aimed at \textit{maximizing} cells quantum efficiency and \textit{minimizing} cells intrinsic layer thickness. 
    Our MOO setup aimed at evaluating several solar cell designs.
    These designs were generated by simulating cells structural components into different compositions by varying transparent conductive oxide layer doping strategy, photodiode materials, back reflector (BR) materials, and roughness of BR layer.
	Of all possible compositions, we selected 15 cell structure designs that are practically feasible and of interest, and for which theoretical models have been validated~\cite{prentice2000coherent,springer2004improved}. 
    Our MOO formulation has a \textit{two-stage strategy} for the optimization of cell structure designs.
    Stage-one applies \textit{non-dominated sorting algorithm--II} (NSGA-II) to produce Pareto-optimal solutions for each cell design. 
    In stage-two, Pareto-fronts of three cell designs selected from stage-one are optimized by using our newly designed algorithm called \textit{multi-objective optimization-immunological algorithm} (OptIA-II). 
    
    Our two-stage strategy improved the baseline quantum efficiency from $0.5999$ to a quantum efficiency of $0.6031$, which is a $0.6845\%$ increase from our baseline on a single solar cell quantum efficiency obtained in~\cite{patane2018enhancing}. 
    Note that a tiny fraction of solar cell efficiency improvement has a proportional impact on solar cell energy production capacity.
    We notice that our methodology can be straightforwardly extended to other solar cell structures, including, e.g., the bifacial one~\cite{sun2018optimization}, where the optimization of optical absorption is a critical issue.

    In summary, the main contributions and observations of this research are as follows:
    \begin{itemize}
    	\item We present a MOO framework for optimal thin-film solar cell structure designs. 
    	\item We optimized, evaluated, and characterized 15 cell designs.
    	\item We present a new algorithm called OptIA-II for MOO of solar cells.
    	\item We show that our two-stage MOO can improve the quantum efficiency of cells and characterize cell designs into clusters concerning to trade-off between cells fabrication cost and cells quantum efficiency. 
    \end{itemize}

    Section~\ref{sec:sloarCell_opt} of this paper explains quantum efficiency of solar cell, solar cell structure and design. Related work on cells structure optimization is discussed in   Section~\ref{sec:related_work}. 
    Formulation of solar optimization as a MOO problem and NSGA-II and OptIA-II algorithms description are presented in Section~\ref{sec:sloarCell_moo}. 
    Section~\ref{sec:exp} describes simulations and experiment setup. Results are discussed in~Section~\ref{sec:results}. 
    Section~\ref{sec:con} outlines conclusions of this work.

	\section{Thin-film silicon solar cells 
		\label{sec:sloarCell_opt}}
	\subsection{Quantum efficiency of solar cells
		\label{sec:Q_e_def}}
	The \textit{quantum efficiency} ($Q_e$) of a solar cell is the ratio of charge carrier produced at the external circuit of the cell (electronic device) to the number of photons received (or absorbed) by the cell.
	There are two ways this quantum efficiency ratio is calculated: 
	(i) external quantum efficiency  and (ii) internal quantum efficiency.

	\textit{External quantum efficiency} is a ratio of the number of photo-generated charge carriers (i.e., electrons) picked up by the cell to the number of photons incident on the cell~\cite{congreve2013external}.

	\textit{Internal quantum efficiency} is the number of electrons (charge carriers) extracted from the cell divided by the number of photons absorbed (trapped) within the cellÕs photoactive layers (i.e., p-i-n-type photodiode absorption layers)~\cite{burkhard2010accounting}. 
	This implies that if all photons absorbed by the photoactive layers results in the generation of charge carriers at the external circuit of the device, then the \textit{internal quantum efficiency} would be 100\%~\cite{slooff2007determining}.
	Hence, maximizing the light absorption would result in maximizing the charge carrier collected at the external circuit~\cite{dewan2009light}.

	This in-fact is thin-film solar cell structure design problem which we discussed in  Section~\ref{sec:sloarCell_design}.
	For this, many candidate cell structures are to be evaluated for achieving maximized \textit{internal quantum efficiency}. 
	We simply call it \textit{quantum efficiency}, $Q_e$. 
	In this work, we compute quantum efficiency, $Q_e$ of a candidate cell  structure according to methods mentioned in Section~\ref{sec:sloarCell_moo}, detail of which is available in~\cite{prentice2000coherent,springer2004improved}.
	
	\subsection{Structure of thin-film silicon solar cells
		\label{sec:thin_film_sloarCell}}
\begin{figure}
		\centering
		\includegraphics[width=0.7\columnwidth]{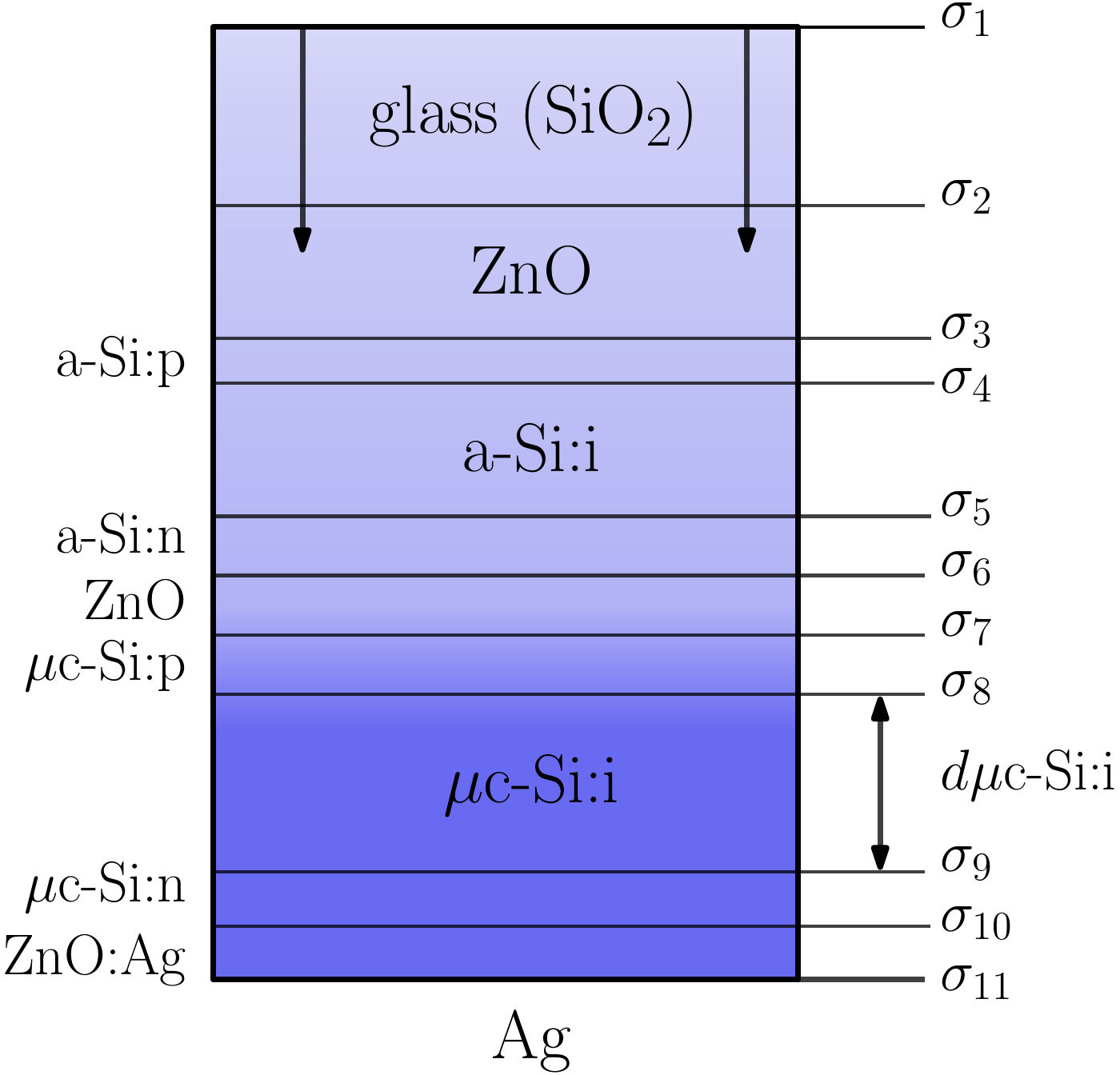}
		\caption{\label{fig:soloar_cell}
			Example of a single thin-film solar cell design, where ZnO is the transparent conductive oxide layer; Ag is the back reflector; $d_{\mu c-Si}$:i  is the thickness of $\mu$c-Si:i layer; and $\sigma_i (i = 1,\ldots,11)$ is the roughness (cells composition parameters that are to be optimized); amorphous silicon (a-Si) and microcrystalline silicon ($\mu$c-Si) are materials in p-i-n photodiodes.}  
	\end{figure}

	This study works on thin-film solar cell composition shown in Fig.~\ref{fig:soloar_cell}.
	The composition of this cell has its \textit{p-i-n-type} doped layers: amorphous silicon (a-Si) and microcrystalline silicon ($\mu$c-Si) separated by a thin ZnO layer (transparent conductive oxide (TCO) layer). 
	This means that the charge careers \textit{p-type} and \textit{n-type} are separated by an \textit{intrinsic layer}. 
	This p-i-n-type photodiode is developed on a layer of composite substance made of SiO$_2$ (glass) and ZnO (zinc oxide). 
	Silver (Ag) or Aluminium (Al) metallic BR supports the cell structure, which is the bottom-most layer of cell in Fig.~\ref{fig:soloar_cell}.

	Glass (SiO$_2$) layer (topmost layer in Fig.~\ref{fig:soloar_cell}) of the cell is the light's entry point. 
	This tandem solar cell structure is benefited from light-harvesting efficiency of its photoactive layers: amorphous silicon (a-Si) and microcrystalline silicon ($\mu$c-Si).  
	Amorphous silicon (a-Si) and microcrystalline silicon ($\mu$c-Si) respectively have $\sim$\SI{1.7}{\electronvolt} and $\sim$\SI{1.1}{\electronvolt} optical band-gaps.
	
    Thin-film technology has made it possible to produce low-cost solar cells. 
    This is mainly due to plasma-assisted chemical vapor deposition technology that enables the production of thin-film solar cells by growing silicon (Si) layers~\cite{klein2004microcrystalline} instead of stacking silicon wafers. 
    Compared with the cost-intensive poly-crystalline Si wafer cutting method where thick poly-crystalline Si wafers are cut and stacked together to form solar cells, the plasma-assisted chemical vapor deposition technology is a highly cost-effective method.

    In-fact, growing Si layers (more generally the absorption layer) also enables the production of solar cells with a smaller thickness (of the order of \SI{}{\micro\meter}) compared to previously possible devices (solar cells) that were thicker by 2-order magnitude.
    However, smaller thickness causes less absorption of light. 
    Therefore, the effective use of light-trapping techniques has become crucial for thin-film solar efficiency.
    The effectiveness of light-trapping techniques relies on thin-film solar cells structure optimization.  
    
    \subsection{Design of solar cells structure 
    	\label{sec:sloarCell_design}}
    When designing and optimizing a solar cell structure, we use two light-trapping methods: light-trapping BR layer and nano-texturing.
	Metals like silver (Ag) maybe used as a BR layer, while alkaline solutions like KOH or NaOH are used for nano-texturing of layer's interfaces. 
	Alkaline solution KOH or NaOH corrodes silicon to form randomly positioned square pyramids.
	The depth of corrosion can be accurately tuned by controlling the temperature and time length of the corrosion process~\cite{rech2002new}. 
	These two methods modify cell geometry (structure).

	Indeed, \textit{cell geometry optimization} helps improve its light-harvesting capacity. 
	The cell geometry (structure) can be optimized by modifying \textit{mesoscale features} (thickness of the different layers) and \textit{nano-scale features} (nano-texturing of the interfaces of the layers).

	The mesoscale features optimization relies on processing time of growing Si layers, and it directly impacts fabrication cost of solar cells. 
	Obviously, mesoscale features optimization is a monotonic function of the layer's thickness (mainly of the intrinsic layer $\mu$c-Si:i)~\cite{shah2004thin}. So is the fabrication cost of solar cells.
	
	Nano-texturing process, on the other hand, produces textured (rough) interfaces in a periodic manner to increase light absorption. The optimization of nano-texturing is essential to make a balance between shading and exposure to light~\cite{mailoa2014textured}. 
	This process can also be performed randomly and roughness (a numeric value measured in \SI{}{\nano\meter}) can be optimized.

	Simultaneous optimization of \textit{mesoscale features} and \textit{nano-scale features} is a trade-off that needs to be met for efficient light-harvesting. 
	On one hand, mesoscale features optimization is sensitive to fabrication costs. On the other hand, nano-texturing feature optimization is sensitive to light absorption capacity. 
	Such a complex optimization problem requires an appropriate numerical approach. 
	This, in fact, is a MOO problem where we optimize the trade-off between the effects of these two cell design features.

	%
	%
	%

    \section{Related work
    	\label{sec:related_work}}
    Thin-film silicon solar cell is relied on light trapping (absorption) techniques to maximize its (internal) quantum efficiency, $Q_e$~\cite{dewan2009light}. 
    Since not all the light entered a cell is absorbed, an optimization of thin-film silicon solar structure design must be performed by varying its structural components for enhancing its light trapping (absorption) capacity~\cite{kabir2012amorphous}.
    Solar cells structural components that can be optimized are
    layers thickness~\cite{patane2018enhancing,kim2014optimization}, 
    layers interface roughness and diffraction grating~\cite{lin2008optimization}, 
    type of materials used in the cell~\cite{jimbo2007cu2znsns4}, 
    and the variations in the BR~\cite{lipovvsek2010modeling,despeisse2011optimization}.
    %

	Numerical simulation~\cite{atwater2011plasmonics} and optical simulation~\cite{prentice2000coherent,springer2004improved} are used for thin-film solar cell structure optimization. 
	These simulations are computationally expansive and solar cell structure design requires evaluation of many designs. 
	Thus, very few works reported to have tackled this problem. 
	Especially, the use of optimization algorithms is rare, leave alone its optimization as a MOO problem.

	An example of genetic algorithm (GA)~\cite{holland1992adaptation}, an optimization algorithm, used for cell structure optimization is available in~\cite{lin2008optimization}, in which the authors aimed at optimizing thin-film cell structure optimization using GA. 
	They vary diffraction gratings (roughness of interface between ZnO:Ag back reflector) of cells. 
	For this, they used periodic rectangular gratings, a four-level rectangular grating, and a random grating resembling a randomly textured surface. 
	GA formulation, therefore, was used to optimize the reflector's geometry with groove height (within range [\SI{0}{\nano\meter}, \SI{800}{\nano\meter}]) and groove period (within rang [\SI{0.5}{\micro\meter}, \SI{5.0}{\micro\meter}]). 
	
	Their purpose was to optimize (to find) grating types related to maximum quantum efficiency, $Q_e$. 
	Compared to their work, our work optimizes each layer's roughness, and our purpose is to optimize photodiode intrinsic layer thickness and composition of cell related to maximum quantum efficiency, $Q_e$. 
	Moreover, our work aims at MOO compared to single-objective optimization mentioned in~\cite{lin2008optimization}.
	
	Contrary to our cell composition optimization approach, some other works in literature aimed at optimizing only a few components of cell structure using single-objective GA. 
	For example, thin-film anti-reflection coatings optimization of solar cell is presented in~\cite{jalali2019genetic} and authors in~\cite{ali2014geometrical} tackles cell structure optimization by optimizing three cell geometrical parameters related to solar cell, %
	namely, number of fingers (balancing shading and surface exposure, i.e., loosely it is the texture of light entry surface) and two values of cells doping profile. 
    
    In our previous work~\cite{patane2018enhancing}, we optimized the trade-off between the cells quantum efficiency, $Q_e$ and the cells structural designs such as the thickness of the cell structure and layers roughness. In our previous work, we evaluated only a few solar cell compositions.  
    However, our current work tackles thin-film solar cell optimization for a variety of TOC materials, BR metals, and TCO doping strategies. 
    We call it a \textit{full scheme} solar cell design and characterization. 
    Here, we aim to evaluate 15 practically feasible cell designs using our two-stage MOO framework.

    \section{Solar cell optimization problem
    	\label{sec:sloarCell_moo}}
    Our work depends on \textit{light trapping} techniques, \textit{plasmonic effects} computation, and \textit{simulation} methods for cells efficiency measurement and cells structure design validation. 
    These methods consist of three parts: 
    (i) the computation of photon's scattering probability for cells layers textured interfaces using Maxwell simulations~\cite{zaidi2002deeply,buddhiraju2017theory};
    (ii) the computation of coherent and scattered photon absorption by a cell structure using a combined method of Photonic Monte Carlo calibration and scattering matrices method; and 
    (iii) the use of optical simulations computational model~\cite{prentice2000coherent,springer2004improved}. 
    Together these methods gave us the measure of solar cells \textit{quantum efficiency} ($Q_e$) (i.e., the ratio between absorbed energy and incident energy) for several parameters like \textit{roughness} of photodiode layers interfaces and \textit{thicknesses} of photodiode intrinsic layer. 
    
    The structure of a thin-film solar cell composition has many layers. Its cell structure optimization is, therefore, subjected to optimization of interface roughness of these layers. 
    We formulated the cell structure optimization problem as a MOO problem (cf. Section~\ref{sec:cell_moo_formulation}), where layers interfaces roughness were the problem's tuneable parameters. 
    The tuning of these parameters governs the optimization of two cost functions related to thin-film solar cells:
    (i) quantum efficiency maximization  and 
    (ii) intrinsic layer thickness minimization.

    \subsection{Solar cell multi-objective optimization formulation
    	\label{sec:cell_moo_formulation}}
    Maximization of solar cell quantum efficiency ($Q_e$)~\cite{prentice2000coherent,springer2004improved} and minimization of microcrystalline  silicon layer thickness ($d_{\mu c-Si}$) are two objectives of the cell structure design.
    The quantum efficiency, $Q_e$ of solar cells is subjected to optimization of their structural factors (parameters), 
    such as (i) roughness of each interface: $\sigma_i, i = 1,\ldots,11$ and (ii) thickness of the microcrystalline silicon layer, $d_{\mu c-Si}$. 
    Therefore, the solar cell optimization problem can be defined as a multi-objective optimization (MOO) problem:\\
    \begin{equation}
    \label{eq:cell_optimization}
    \begin{aligned}
    &\text{maximize} \quad Q_e\\
    &\text{minimize} \quad d_{\mu c-Si}\\
    &\text{subject to} 	\left\lbrace
    \begin{array}{l} 
    \sigma_1 = 0 \\
    0 \le \sigma_2 \le 0.3 \bar{d_2} \\
    0 \le \sigma_i \le 0.2 \bar{d_i} \quad i = 3,\ldots,11  \\
    0.7\bar{d}_{\mu c - Si} \le d_{\mu c - Si} \le 1.3\bar{d}_{\mu c - Si}\\
    \end{array}\right.
    \end{aligned}
    \end{equation}
    where $Q_e$ is the average overall quantum efficiency of a cell in an ideal charge collection condition (i.e., 100 \% collection efficiency independently collected on the frequency) calculated as the average absorption in the two intrinsic layers
    $\mu$c-Si:i and a-Si:i. 
    In Eq.~\eqref{eq:cell_optimization}, $d_{\mu c-Si}$ is the thickness of intrinsic ($\mu$c-Si) layer, and $ \bar{d_i}$ is the thickness pertaining to $i$-th material of the reference cell (cf. Table~\ref{tab:refernce_cell}).

    The cell structure optimization concerning intrinsic ($\mu$c-Si) layer's thickness, $d_{\mu c-Si}$ minimization is directly related to cells fabrication cost. 
    However, solar cell optimization problem imposes a minimum and maximum thickness constraint on intrinsic ($\mu$c-Si) layers in its MOO treatment to Eq.~\eqref{eq:cell_optimization}. 
    The constraint in Eq.~\eqref{eq:cell_optimization} maintains the light scattering weight where intrinsic ($\mu$c-Si) layer is thinner than the ZnO layer's roughness.
    Moreover, the approximation made by modelling approach refereed in Section~\ref{sec:thin_film_sloarCell} for cell structure optimization and validation holds true only if the constraints in Eq.~\eqref{eq:cell_optimization} are met~\cite{ward1988optical}.
    \begin{table}
        \centering
        \renewcommand{\arraystretch}{1.2}
        \setlength{\tabcolsep}{20pt}
        \caption{\label{tab:refernce_cell} 
        	Reference values of layers thickness of layer's interface roughness. The Materials and structure in Table~\ref{tab:refernce_cell} is a reference cell belongs to the solar cell shown in Fig.~\ref{fig:soloar_cell} that has zinc oxide-based transparent conductive oxide layer and silver as a back reflector and amorphous silicon (a-Si) and microcrystalline silicon ($\mu$c-Si) as p-i-n-type photodiodes layers.
        	A different set of ``Materials'' produces its respective  \textit{reference cell} for the ``Thickness (\SI{}{\nm})'' and ``Roughness (\SI{}{\nm})'' values as mentioned in this Table.}
        \begin{tabular}{lll}
            \toprule
            Material & Thickness (\SI{}{\nm}) & Roughness (\SI{}{\nm})\\
            \midrule
            Glass (SiO$_2$) & $3.5\times10^6$ & $\sigma=0$\\
            Zinc-Oxide (ZnO)         & $600$  & $\sigma=120$ \\
            a-Si:p      & $20$   & $\sigma=2$   \\
            a-Si:i      & $300$  & $\sigma=30$  \\
            a-Si:n      & $20$   & $\sigma=2$   \\
            ZnO         & $20$   & $\sigma=2$   \\
            $\mu$c-Si:p & $20$   & $\sigma =2$  \\
            $\mu$c-Si:i & $1.7\times10^3$  & $\sigma=170$\\
            $\mu$c-Si:n & $20$   & $\sigma =2$  \\
            ZnO:Ag      & $20$   & $\sigma =2 $ \\
            Ag          & $150$  & $\sigma =15 $\\
            \bottomrule
        \end{tabular}
    \end{table}

	\subsection{Multi-objective optimization algorithms}
	A multi-objective optimization (MOO) algorithm optimizes two or more objectives simultaneously as per:
    \begin{equation}
     F(\textbf{x}) \equiv  (f_1(\textbf{x}), \ldots, f_m(\textbf{x})), \mbox{ i.e., } F: \mathbb{R}^n \rightarrow \mathbb{R}^m \mbox{ for } m \ge 2    
    \end{equation}
    such that no one objective of the problem can be improved without a simultaneous detriment to at least one of the other objectives. 
    Each $ f_k(\textbf{x})$ is a scalar objective, and MOO problem optimizes an objective vector $F(\textbf{x})$, where $\textbf{x} \in \mathbb{R}^n$ is its feasible solution obtained at the final population on meeting termination criteria like max number of function evaluations.

    More specifically, a MOO algorithm produces a set of non-dominated solutions $\{\textbf{x}_1, \textbf{x}_2,\ldots,\textbf{x}_\lambda\}$, also known as \textit{Pareto-optimal front} or \textit{Pareto-optimal solutions set}~\cite{deb2002fast}, which is defined as follows. 
    
    A solution $ \textbf{x}_i $ \textit{dominates} other solution $ \textbf{x}_j $ if for all objectives $ f_k = 1, 2,\ldots m $, $f_k(\textbf{x}_i) \preccurlyeq f_k(\textbf{x}_j)$, where $\preccurlyeq$ should be read as ``better of.'' 
    On the contrary, a solution $ \textbf{x}_i $ is \textit{non-dominated} if for at least one objective $f_k(\textbf{x}_i) \preccurlyeq f_k(\textbf{x}_j)$ does not hold.
    A set of such non-dominated solutions is called the Pareto-optimal solutions set.    
    
    \indent
    The solar cell structure optimization problem optimizes the objective vector:
    \begin{equation}
        F = (Q_e(\boldsymbol\sigma), d_{\mu c-Si}(\boldsymbol\sigma)),
    \end{equation}
    where $Q_e$ and $d_{\mu c-Si}$ are subjected to optimization of parameters (roughness of cell structure)  $ \boldsymbol\sigma = \left\langle \sigma_2, \ldots, \sigma_{11}\right\rangle$. 
    We report three select solutions from the Pareto-front of each cell design: 
    (i) the solution maximizing the quantum efficiency, i.e., $Q_e(\boldsymbol\sigma)^*$;
    (ii) the solution minimizing the thickness of the microcrystalline silicon layer ($d_{\mu c-Si}$), i.e., $d_{\mu c-Si}(\boldsymbol\sigma)^*$; and 
    (iii) the solution closest to the Utopian point, i.e.,  $U(\boldsymbol\sigma)^*$ which represents the solution that  contributes highest to both objectives with respect to the reference cell mentioned in Table~\ref{tab:refernce_cell}.

	We used a \textit{two-stage approach} to perform MOO of solar cell. 
	In the first stage, we performed non-dominated sorting genetic algorithm-II (NSGA-II) based optimization of the objective $ F=(Q_e(\boldsymbol\sigma), d_{\mu c-Si}(\boldsymbol\sigma)) $. 
	In the second stage, we apply our multi-objective optimization-immunological algorithm (OptIA-II) to enhance the Pareto-front obtained in the first stage. 
	We briefly describe NSGA-II and OptIA-II as follows.  

	\subsubsection{Non-dominated sorting genetic algorithm-II (NSGA-II) 
		\label{sec:nsga_II}}
	NSGA-II~\cite{deb2002fast} is an evolutionary algorithm inspired by the principle of natural selection and has been found effective in real word applications (e.g.,~\cite{umeton2011design}). 
	NSGA-II uses non dominated sorting, elitism, and crowding distances strategies to guide an initial population $P = (\boldsymbol\sigma_1, \ldots, \boldsymbol\sigma_D)$ of $ D $ solution vectors through a predefined number of steps to a final population while simultaneously optimizing trade-off of multiple objectives. 
	Operators like crossover and mutation generate new solutions in each generation. This offspring generated from the current population, is then joined to their parents to constitute an enlarged set of vectors that are then sorted to choose the population that will be passed to the next generation of the algorithm.
	To perform this selection, in each generation, each solution of population $P$ is assigned a rank and a crowding distance value.

	A \textit{fast non-dominated sorting} process sorts $ D $ candidate solutions into several sets (called fronts) of non-dominated solutions: $F_1,F_2, \ldots,F_s$ such that the front $ F_1 $ contains all the non-dominated candidate solutions of population $P$.
	This means, no one solution in $F_1$ is dominated by any other solutions.

	Considering all remaining solutions except the ones already in $F_1$, it is possible to identify a new front $F_2$ which contains all the next non-dominated solutions of $P$. 
	Subsequently, front $F_3$ and other fronts are sorted until no solution is left to be sorted into a front.

	Once solutions are sorted into several fronts, a rank is assigned to each candidate solution such that those in front $F_1$ have rank 1, those in front  $F_2$ have rank 2, and those in front  $F_3$ have rank 3. This continues until all solutions are assigned a rank.
	
	The rank of the solutions is then used as the first parameter considered when constructing the population for the next generation.
	
	The \textit{crowding distance} of a candidate solution $ \boldsymbol\sigma  $ is the estimation of the density of solutions around it. 
	The distance assigned to $ \boldsymbol\sigma  $ is the sum of its distance from the two points found on either side of it in the objective space.
	Crowding distance helps NSGA-II maintain diversity in the population while propagating population, $P$ from generation to generation, by promoting the solutions with a higher distance from the other points within the same front.

	\subsubsection{Multi-objective  Immunological algorithm (OptIA-II) 
		\label{sec:opita_ii}}
	We designed a novel multi-objective optimization  algorithm called OptIA-II. 
	Our OptIA-II is a clonal selection algorithm \cite{Nicosia:PhD05,Nicosia07}.
	There are three immunological theories: immune networks, negative selection, and clonal selection. 
	Of which, OptIA-II follows \textit{clonal selection} theory \cite{Nicosia:PhD05}.
	OptIA-II works on the following four principles: (i) cloning, (ii) inversely proportional hypermutation, (iii) hypermacromutation, and (iv) aging.
	
	\paragraph{Cloning operator.}
	The \textit{cloning operator} clones each candidate solution so that these clones can be modified using the other operators to explore the neighbourhood in the search space \cite{Nicosia:PhD05,Nicosia05} while keeping the original parents points of the population.
	
	\paragraph{Hypermutation and hypermacromutation operators.}
	Hypermutation and hypermacromutation operators \cite{Nicosia:PhD05} perturb each candidate solution using an inversely proportional law where a candidate solution with a better objective rank has a lower number of its genes mutated, and a candidate solution with a poorer objective rank has a higher number of its genes mutated.
	Hypermutation and hypermacromutation differ  by the number gens they mutate for best and worse solutions. Hypermacromutation mutates relatively low number of genes compared to  hypermutation. For example, if hypermacromutation is expected to mutate about $5\%$ and $15\%$ of the genes of the best and worst solutions respectively, the hypermutation is expected to mutate about $37\%$ of the genes of the best solution and $100\%$ of the genes of the worst solution. However, the hypermacromutation will likely apply more significant changes in the genes it mutates.
	
	The formal definitions of hypermutation and hypermacromutation operators are as follows:
	Let $n$ be the number of genes a candidate solution $\boldsymbol\sigma$ has. 
	Then, we have $n =10$ genes of a candidate solution subjected to mutation in this work, i.e., $\boldsymbol\sigma = \sigma_2,\sigma_3,\ldots,\sigma_{11}$ representing layers interface roughness. 
	For convenience, we re-index solution vector as  $\boldsymbol\sigma = \sigma_1,\sigma_2,\ldots,\sigma_{10}$.
	

    A threshold $\alpha$ is defined for each solution and used to determine whether its genes will be mutated by the {\it hypermutation operator}, i.e. it expresses the probability with which the mutation of a gene occurs.
    
    Thus, for a hypermutation potential $\rho_1$, we compute $\alpha$ as: 
    \begin{equation}
        \label{eq:mutation_rate}
        \alpha=e^{-\rho_1\hat{F}(\boldsymbol\sigma)},
    \end{equation}
    
    where $\hat{F}(\boldsymbol\sigma)$ indicates normalized objective rank between $[0,1]$ of the candidate solution $\boldsymbol\sigma$, where $1$ is the best solution and $0$ is the worse solution. Once $\alpha$ is defined, the expected number of genes mutated by the operator is approximately equal to $\lfloor \alpha \cdot n + 0.5 \rfloor$.
    
    In practice, for each gene a random variable is then drawn from a uniform distribution $\mathcal{U}(0,1)$ and then compared to $\alpha$, if the variable is below the threshold the hypermutation of that gene is performed as:
    	\begin{equation}
	    \label{eq:mutation_operation}
	    \sigma_i = \sigma_i \cdot e^{(\tau_1\cdot r)+(\tau_2\cdot s)},
	    \end{equation}
    where $\tau_1, \tau_2 < 1$ are two parameters dependent on the population size $D$ and defined as	$\tau_1=1/\sqrt{2\sqrt{D}},
    \tau_2=1/\sqrt{2D}$ and $r,s \sim \mathcal{N}(0,1)$ are two variables drawn from a standard normal distribution.
    
    \begin{algorithm}
        \caption{Hypermutation operator}
        \label{algo:algo_hyperm}
        \begin{algorithmic}
            \Function{Hypermutation}{$(P_{\text{clone}},\rho_1)$ }
            \For{$j=(1, \ldots, D_{clone}$)}
        	\State ${\boldsymbol\sigma} = P_{clone_j}$
        	\State  $\alpha = e^{\rho_1*\hat{F}(\boldsymbol\sigma)}$ \Comment{Threshold for mutation is computed}
        	\State $r, s \sim \mathcal{N}(0,1)$
        	\State $u \sim \mathcal{U}(0,1)$
            \For{$i=1, \ldots, n$} \Comment{For each gene in ${\boldsymbol\sigma}$}
              \If{$u < \alpha$}
              \While{$\#try \leq 10$} \Comment{Try 10 time to achieve a feasible mutation}
              \State $\mathbf{\sigma}_i = \mathbf{\sigma}_i \cdot e^{(\tau_1 \cdot r) + (\tau_2 \cdot s)}$ \Comment{Mutation of the gene}
              \EndWhile
    	      \EndIf
	        \EndFor
	        \State $P_{clone_j} = {\boldsymbol\sigma}$
	        \EndFor
	        \Return $P_{clone}$
	        \EndFunction
    	 \end{algorithmic}
    \end{algorithm}
    
    On contrary to hypermutation, 
    in the {\it hypermacromutation operator} the threshold $\alpha$ to determine the genes of each candidate solution $\boldsymbol\sigma$ to be mutated, for a hypermacromutation potential $\rho_2 $, is computed as:
	\begin{equation}
	    \label{eq:macromutation_rate}
	    \alpha = \frac{e^{-\hat{F}(\boldsymbol\sigma)}}{\rho_2}. 
	\end{equation}
	Similarly to the hypermutation case, a random variable drawn for each gene $i$ of the solution triggers the hypermacromutation if lower than $\alpha$.
	This mutation is performed by using a convex operator which randomly selects another gene $\sigma_j$ for $i\ne j$ and by applying the following operation:
	\begin{equation}
	    \label{eq:convex_mutation_operation}
	    \sigma_i = (1 - \beta) \sigma_i + \beta \left(l_i+\frac{\sigma_j-l_j}{u_j-l_j}(u_i-l_i)\right),
	\end{equation}
	where $\beta \in [0,1]$ is drawn from a uniform distribution, and $l$ and $u$ are the lower and upper bounds defined for each gene, respectively. Hence, the operator applies a convex mutation based on the value of the gene $\sigma_i$ to be mutated and the value of another gene $\sigma_j$ that is normalized and transferred in the feasible interval of $\sigma_i$.

	 \begin{algorithm}
        \caption{Hypermacromutation operator}
        \label{algo:algo_hypermacro}
        \begin{algorithmic}
            \Function{Hypermacromutation}{$(P_{\text{clone}},\rho_2)$}
            \For{$j=(1, \ldots, D_{clone}$)}
        	\State ${\boldsymbol\sigma}$ = $P_{clone_j}$
        	\State  $\alpha = \frac{e^{-\hat{F}(\boldsymbol\sigma)}}{\rho_2}$ \Comment{Threshold for mutation is computed}
        	\State $j = randomInt([1,n]), j \neq i$ \Comment{Choose a random gene for the convex mutation}
        	\State $u \sim \mathcal{U}(0,1)$
            \For{$i=1, \ldots, n$} \Comment{For each gene in $\mathbf{\sigma}$}
              \If{$u < \alpha$}
              \State $\beta \sim \mathcal{U}(0,1)$
              \State $\sigma_i = (1 - \beta) \sigma_i + \beta \left(l_i+\frac{\sigma_j-l_j}{u_j-l_j}(u_i-l_i)\right)$ \Comment{Convex mutation is applied}
    	      \EndIf
	        \EndFor
	        \State $P_{clone_j} = {\boldsymbol\sigma}$
	        \EndFor
	        \Return $P_{clone}$
	        \EndFunction
    	 \end{algorithmic}
    \end{algorithm}
    
    A mentioned, the two mutation operators have similarities and differences; first, the definition of the probability of the mutation $\alpha$ involves different parameters such as the two mutation potentials $\rho_1, \rho_2$ and, in the case of hypermutation, $\tau_1, \tau_2$ that are dependent on the population size only.
    The other notable difference is how the mutation is performed; hypermutation is a variation of the corresponding gene based on its value and an exponential function that maintains the new value close to the original one due to the definition of $\tau_1, \tau_2$, having $D = 100$. The convex operator in the hypermacromutation makes the mutated value more likely to be far from the original gene, also depending on its normalized difference with the other randomly chosen gene in the solution.
    
    That is, hypermacromutation aims to produce a more diverse candidate solution than the original one, improving the diversity across the population, while the hypermutation keep the exploration more within the close range of the existing genes.
    
    In our case, posing $\rho_1 = 1, \rho_2 = 7$ makes the chance of a hypermutation of a gene more likely that a hypermacromutation. The presence of the normalized rank of solutions $\hat{F}(\boldsymbol\sigma)$ in the definition of the probabilities ensures the inversely proportional law, making both mutations more likely to occur within the worst solutions of the current population.
    
    At every step of the procedures in which new genes are generated, they are maintained only if feasible as defined by the constraints of the problem and discarded otherwise.
	
	\paragraph{Aging operator.}
	The \textit{aging operator} eliminates old candidate solutions from the current population to introduce diversity and to avoid local minima \cite{Nicosia:PhD05}.
	Aging operator depends on the number of iterations a solution can survive in the population. For example, if we set an \textit{age limit} of 50 iterations for solutions in a population. Then, the solution will be dropped from the population. However, the best solution in the population is waived from this elimination (\textit{elitism}).
	
	Moreover, the best \textit{dead} solutions are retained in an archive to be considered again if during a generation the number of feasible and alive solutions found after the mutations and aging goes below the size of the population.
	
	\paragraph{Pareto-front-selection operator.}
	After aging operator has eliminated older solutions from the current population, a \textit{Pareto-front-selection operator} is performed to assign a rank to each solution of a combined set of solutions of both current population (without aged solution) and current offspring solutions.
	The Pareto-front-selection operation in OptIA-II is a fast non-dominated sorting of population, $P$, which is like fast non-dominated sorting of NSGA-II described in Section~\ref{sec:nsga_II}.
	
    \paragraph{OptIA-II optimization cycle.}
    OptIA-II outlined in Algorithm~\ref{algo:algo_optia} starts with an initial population $ P $ of $ D $ candidate solutions. 
    Each member of the population is a candidate solution for the given optimization problem. 
    Cloning, hypermutation, hypermacromutation, elitism, and aging operators of  OptIA-II algorithm guides an initial population $P = (\boldsymbol\sigma_1, \ldots, \boldsymbol\sigma_D)$ of $ D $ solution vectors through a predefined number of steps to a final population while simultaneously optimizing trade-offs of multiple objectives.

    In main optimization cycle of OptIA-II, the cloning operator clones (duplicates) each candidate solution $D'$ times. 
    On these solutions, hypermutation and hypermacromutation operators are applied with respective mutation potentials $\rho_1$ and $\rho_2$ to mutate randomly chosen candidate solutions. 
    
    After fitness of each solution is evaluated, an aging operator discards solutions that have reached a predefined age limit (the maximum number of steps $\tau$ that a solution can survive in the population). 
    However, elitism preserves the best solution irrespective of its age. 
    Finally, remaining solutions of the population are assigned rank and sorted as per Pareto-front-selection operation \cite{Nicosia12}.
    \begin{algorithm}
        \caption{Multi-objective optimization-immunological algorithm (OptIA-II)}
        \label{algo:algo_optia}
        \begin{algorithmic}
            \Procedure{OptIA-II Algorithm}{$ P$, $D'$, $\tau$, $\rho_1$, $\rho_2$, $T_{max}$, $F$}
            \State $\rho_1 \leftarrow $ hypermutation potential
            \State $\rho_2 \leftarrow $ hypermacromutation potential
            \State $\tau \leftarrow $ maximum age
            \State $t = 0$ \Comment{generation initialization}
            \State $P_t = (\boldsymbol\sigma_1,\ldots, \boldsymbol\sigma_D)$ \Comment{a population $P$ of $D$ solutions initialized uniformly at random}
            \State $\boldsymbol\sigma_i^{age} = 0$ \Comment{at iteration $0$ each candidate solution $i$ has age 0}
            \State $D_{\text{clone}} = D.D' $ \Comment{number of clones $D_{\text{clone}}$ for cloning factor $ D' $}
            \State $P_t \leftarrow $ \textsc{Fitness Evaluation} $ F(P_t)$ \Comment{fitness evaluation for $F = (Q_e(\boldsymbol\sigma), d_{\mu c-Si}(\boldsymbol\sigma))$}
            \State $F_{\text{eval}} = D $ \Comment{initial total number of fitness evaluations set to $D$}
            \While{$F_{\text{eval}} < $  $T_{max}$}
            \State $P_{\text{clone}} \leftarrow $ \textsc{Cloning}$(P_t, D_{\text{clone}})$  
            \State $P_{\text{hyp}} \leftarrow $ \textsc{Hypermutation}$(P_{\text{clone}},\rho_1)$ 
            \State $P_{\text{macro}} \leftarrow $ \textsc{Hypermacromutation}$(P_{\text{clone}},\rho_2)$
            \State $P_{\text{hyp}} \cup P_{\text{macro}} \leftarrow $ \textsc{Fitness Evaluation} $ F\left(P_{\text{hyp}} \cup P_{\text{macro}}\right)$
            \State $F_{\text{eval}} = F_{\text{eval}} + 2.D_{\text{clone}} $
            \State $P_{\text{age}} \leftarrow $ \textsc{Aging}$(P_t, P_{\text{hyp}} \cup P_{\text{macro}}, \tau)$
            \State $P_{t+1} \leftarrow $ \textsc{Pareto-Front-Selection}$(P_{\text{age}})$
            \State $t = t + 1$
            \EndWhile
            \Return $P_{T_{max}}$
            \EndProcedure 
        \end{algorithmic}
    \end{algorithm}

	\section{Simulations and experiment setting 
		\label{sec:exp}}
    
    \subsection{Solar cell designs}
    
	We present a comprehensive and exhaustive simulation strategy for solar cell optimization in this work. 
	We performed a total of 15 different simulations in our experimental framework. 
	These simulation cases are shown in Table~\ref{tab:simulation_detail}. 
	We aim to investigate the structural validity and practical feasibility of these cases.

	Compared to our previous work~\cite{patane2018enhancing}, where four levels of dosage of TCO and two varieties of BR roughness were used, we explore more comprehensive solar cell design variations in this work.  
	We investigated the following \textit{design variations}: 
	\begin{itemize}
		\item four-levels of TCO dosage---\textit{optimally} doped (resistance less than \SI{1}{\meter\ohm}$\times$\SI{}{\centi\meter}), \textit{normally} doped (resistance of the order of \SI{1}{\meter\ohm}$\times$\SI{}{\centi\meter}), \textit{lowly} doped (resistance larger than \SI{100}{\meter\ohm}$\times$\SI{}{\centi\meter}), and \textit{not} doped; 
		\item two types of TCO materials-- \textit{zinc-oxide} (ZnO) and \textit{tin-oxide} (SnO2); 
		\item two varieties of BR roughness--\textit{smooth} and \textit{rough} surfaces; and 
		\item two types of back reflector--metals \textit{silver} (Ag) and \textit{aluminium} (Al). 
	\end{itemize}

    Such a comprehensive and computationally expensive setup was explored to investigate solar cells structure designs and materials that could maximize the quantum efficiency, $Q_e$ and minimize intrinsic layer thickness, $d_{\mu c-Si}(\boldsymbol\sigma)$. 
    Moreover, we hope that it will produce a full characterization of solar cell designs and associate the designs with their quantum efficiency and fabrication cost trade-off.

    The solar cell structure in Fig.~\ref{fig:soloar_cell} is a layer-wise composition.
    The layers are designed by varying the mentioned four categories of variations (cf. Table~\ref{tab:simulation_detail}). 
    Each design, therefore, requires approximation of its layer interface roughness $ \boldsymbol{\sigma} $ that maximizes its quantum efficiency and minimizes its fabrication cost. 
	\begin{table}
        \centering
        \renewcommand{\arraystretch}{1.2}
        \setlength{\tabcolsep}{5pt}
        \caption{Solar cell structure design compositions for simulation}
        \label{tab:simulation_detail}
        \begin{tabular}{ll}
            \toprule
            Simulation & TCO, back reflector, and doping variations \\
            \midrule
            & \textit{Cost intensive TOC and back reflector metal use}\\
            sAg-oZnO & Smooth back reflector Ag, Optimally Doped ZnO (the most cost-intensive) \\
            sAg-nZnO & Smooth back reflector Ag, Normally Doped ZnO\\
            sAg-lZnO & Smooth back reflector Ag, Lowly Doped ZnO \\
            sAg-ntZnO & Smooth back reflector Ag, Not Doped ZnO \\[3pt]
            
            & \textit{Cost intensive back reflector roughness}\\
            rAg-oZnO & Rough back reflector Ag, Optimally Doped ZnO\\
            rAg-nZnO & Rough back reflector Ag, Normally Doped ZnO\\
            rAg-lZnO & Rough back reflector Ag, Lowly Doped ZnO \\
            rAg-ntZnO & Rough back reflector Ag, Not Doped ZnO\\[3pt]
            
            & \textit{Cost-effective back reflector metal use} \\
            sAl-oZnO & Smooth back reflector Al, Optimally Doped ZnO \\
            sAl-nZnO & Smooth back reflector Al, Normally Doped ZnO\\
            sAl-lZnO & Smooth back reflector Al, Lowly Doped ZnO\\
            sAl-ntZnO & Smooth back reflector Al, Not Doped ZnO\\[3pt]
            
            & \textit{Cost-effective TOC use}\\		
            sAg-SnO2 & Smooth back reflector Ag, SnO2\\
            rAg-SnO2 & Rough back reflector Ag, SnO2\\
            sAl-SnO2 & Smooth back reflector Al, SnO2 (the least cost-intensive)\\
            \bottomrule
        \end{tabular}
    \end{table}

    \subsection{Two-stage multi-objective optimization strategy}
    In our two-stage optimization strategy, the \textit{first stage} uses NSGA-II algorithm.
    NSGA-II aims at maximizing quantum efficiency and minimizing intrinsic layer thickness simultaneously. 
	NSGA-II based optimization was performed on all 15 cell designs. 
	The simulation results of each design's feasible solutions were collected and compared with their respective reference cells.

    For NSGA-II based MOO, population size was $ P = 100 $,  and a maximum of 200 generation was used as the stopping criterion. 
    We used \textit{simulated binary crossover} (SBX) with crossover probability $ pc = 0.9 $. 
    For mutation, \textit{polynomial mutation} with mutation probability $ pm = 1 / (\text{number of decision variables}) $ was used. 
    The parent selection method was \textit{binary tournament selection}.

    The \textit{second stage} was the fine-tuning stage. 
    Here, we apply our OptIA-II algorithm to the results of NSGA-II. 
    In this stage, OptIA-II aimed at improving select best Pareto-fronts obtained using the NSGA-II algorithm. 
    Three best designs obtained in the first stage were optimized in the second stage.   
    
    For OptIA-II, the population size was the same as the Pareto-front population size of the NSGA-II algorithm. 
    The {cloning factor} $ D' $ was $ 2 $, the {aging factor} $ \tau$ (maximum age) was $ 50 $ iterations, {hypermutation potential}l $ \rho_1 $ was 1, {hypermacromutation potential} $ \rho_2 $ was 7. 
    The stopping criteria $ T_{max} $ was $10^4  $ fitness function evaluations.

	\section{Results and discussion
		\label{sec:results}}
    In our previous work~\cite{patane2018enhancing}, a solar cell design ``highly doped ZnO and smooth Ag as the BR'' on optimization using a single-objective algorithm showed a 5.88\% improvement with respect to the reference cell (i.e., $Q_e = 0.604169$ and $d_{\mu c-Si}=2210$). 
    However, for this cell design, no constraint mentioned in Eq.~\eqref{eq:cell_optimization} was applied. 
    Thus, parameter values $ \boldsymbol{\sigma} $ were very high. 
    This made the obtained design fragile and unstable.

    Using single objective OptIA without violating constrained in Eq.~\eqref{eq:cell_optimization}, it was possible to obtain a solution with $Q_e = 0.5928$ and $d_{\mu c-Si}=1814.45;$ while NSGA-II based MOO produced a $5.083$\% improvement on absorption ($Q_e = 0.599967$ and $d_{\mu c-Si}=2209.97876$) with respect to the reference cell design (cf. Table~\ref{tab:refernce_cell}).

    In addition, another solar cell design ``low doped ZnO and rough BR Ag'' produced a $Q_e = 0.6$ and $d_{\mu c-Si}=2100$ on NSGA-II-based optimization. 
    For ``optimally doped ZnO and smooth BR Ag,'' our previous work produced a $Q_e = 0.593$ and $d_{\mu c-Si}=2110$ using single objective OptIA as its best results~\cite{patane2018enhancing}.

    In this research, we use the max quantum efficiency, $Q_e = 0.6$ obtained in~\cite{patane2018enhancing} using NSGA-II for cell design ``low doped ZnO and rough BR Ag'' as our baseline accuracy. 
    Moreover, we aim to maximize quantum efficiency, $Q_e$ as much as possible while maintaining the constraint laid down in Eq.~\eqref{eq:cell_optimization}. 
    We hope to obtain stable cell designs from our two-stage MOO framework with quantum efficiency, $Q_e > 0.6$ for their intrinsic layer thickness within the range $[1190, 2210]$.

	\subsection{Stage 1: Multi-objective optimization using NSGA-II}
\begin{figure}
        \centering
        \includegraphics[width=\columnwidth]{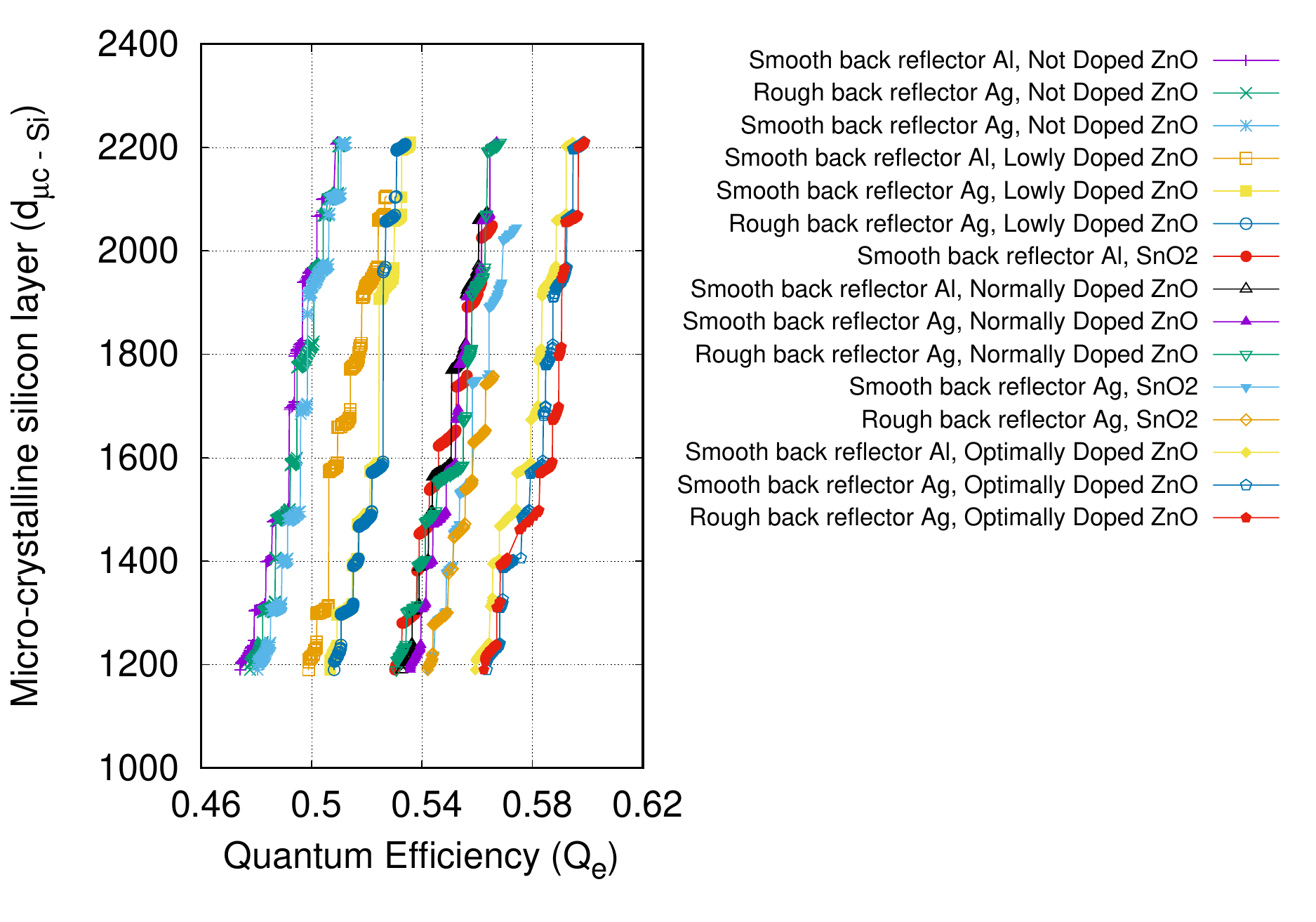}
        \caption{Pareto-front approximations obtained by using NSGA-II for the multi-objective optimization (minimization of Micro-crystal silicon layer $d_{\mu c} - S_i$ and maximization of Quantum efficiency $Q_e$) of the optical model for tandem-thin-film silicon solar cell with the ''full parameterized'' model for all 15 solar cell simulation variations. Notice that the Pareto-front of all simulations is coloured from  the lower Quantum efficiency $Q_e$ value to higher (apparently best) the best Quantum efficiency $Q_e$. Design with ``Rough back reflector Ag and Optimal Doped ZnO version is indicted in red line with red pentagon marker appear, which also indicate the best Pareto-front among all designs.}
        \label{fig:simulation_all_results}
    \end{figure}

    A summarized result of NSGA-II based MOO of the solar cell optimization on all 15 designs is shown in Fig.~\ref{fig:simulation_all_results}. 
    In stage-one, NSAG-II offers best quantum efficiency, $Q_e = 0.6$ on cell design configuration ``optimally doped ZnO and smooth BR Ag.''  
    This result is the same as the baseline in our previous experiment. 
    This motivated us to fine-tune some select best results of stage-one in stage-two.

    However, in our current frameworks, we hope not only for maximizing cells quantum efficiency, $Q_e$, but we hope to have a \textit{full characterization} of cell designs and approximate their design parameters, which our stage-one provides successfully.
	Fig.~\ref{fig:simulation_all_results} offers this characterization by summarizing Pareto-fronts of all 15 designs.

    In Fig~\ref{fig:simulation_all_results}, quantum efficiency, $Q_e$ maximization trait is plotted on x-axis against minimization trait of microcrystalline silicon layer thickness, $d_{\mu c-Si}$ plotted on y-axis. 
    We notice that this summarized plot of solar cell designs generates clusters that are linked to different design characteristics. 
    Here, four clear clusters (four sets of cell designs) can be seen in Fig.~\ref{fig:simulation_all_results}.   
    We characterize each cluster as follows:

    \paragraph{Lowest cost-intensive and lowest quantum efficient cluster.}
    The solar cell designs that are \textit{not-doped} constitute this cluster. 
    This cluster appears in the leftmost part of Fig.~\ref{fig:simulation_all_results} (first three labels listed on the right panel). 
    The quantum efficiency, $Q_e$ of \textit{not-doped} cell designs vary between $0.47$ and $0.51$. 
    These quantum efficiencies are obviously less than baseline quantum efficiency, $Q_e = 0.6$. 
    This makes this cluster the least quantum efficient cluster. 

\begin{figure}[h!]
    	\centering
    	\subfigure[]{
    		\includegraphics[width=0.32\columnwidth]{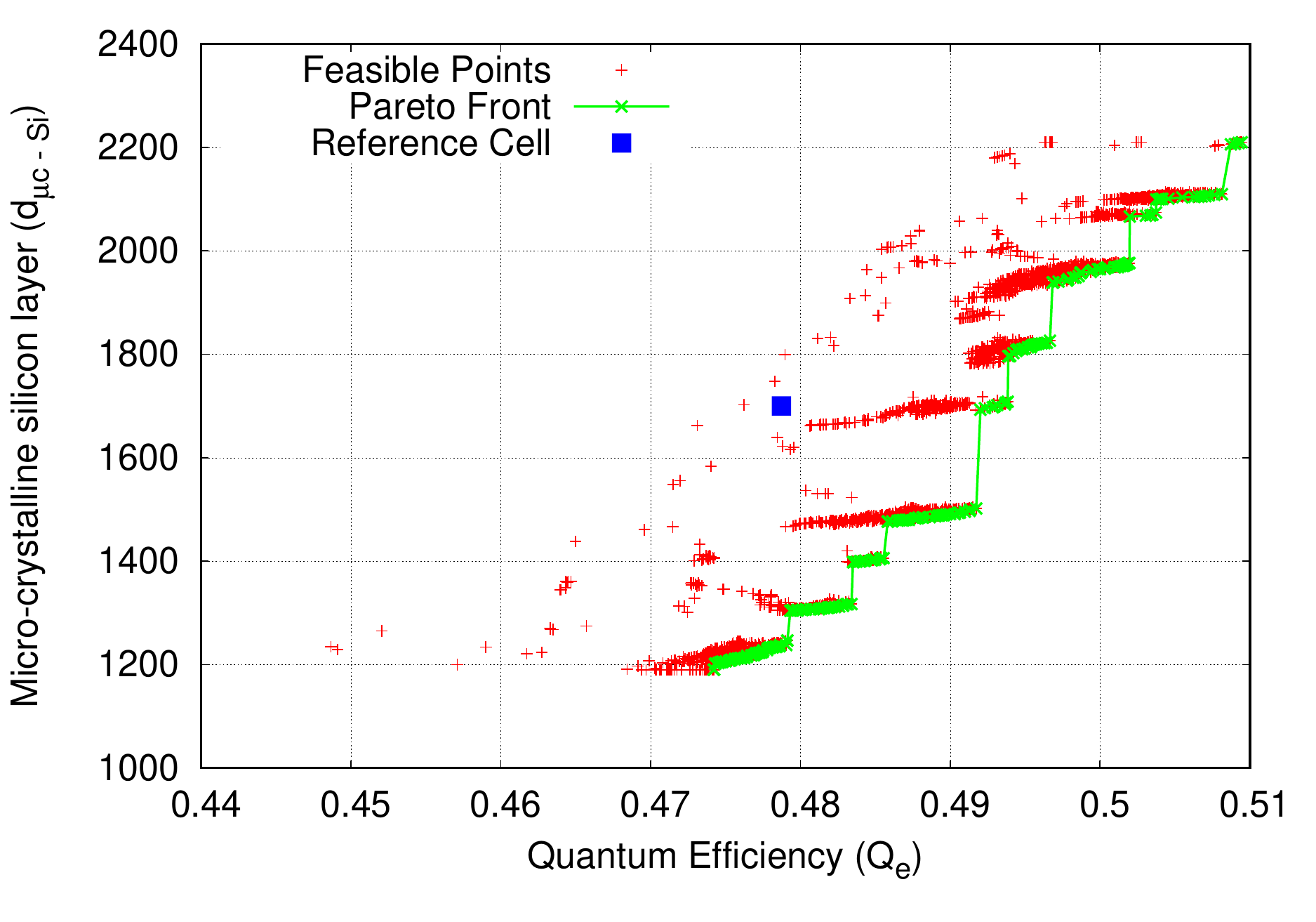}\label{fig:sAl_ntZnO}}
    	\subfigure[]{\includegraphics[width=0.32\columnwidth]{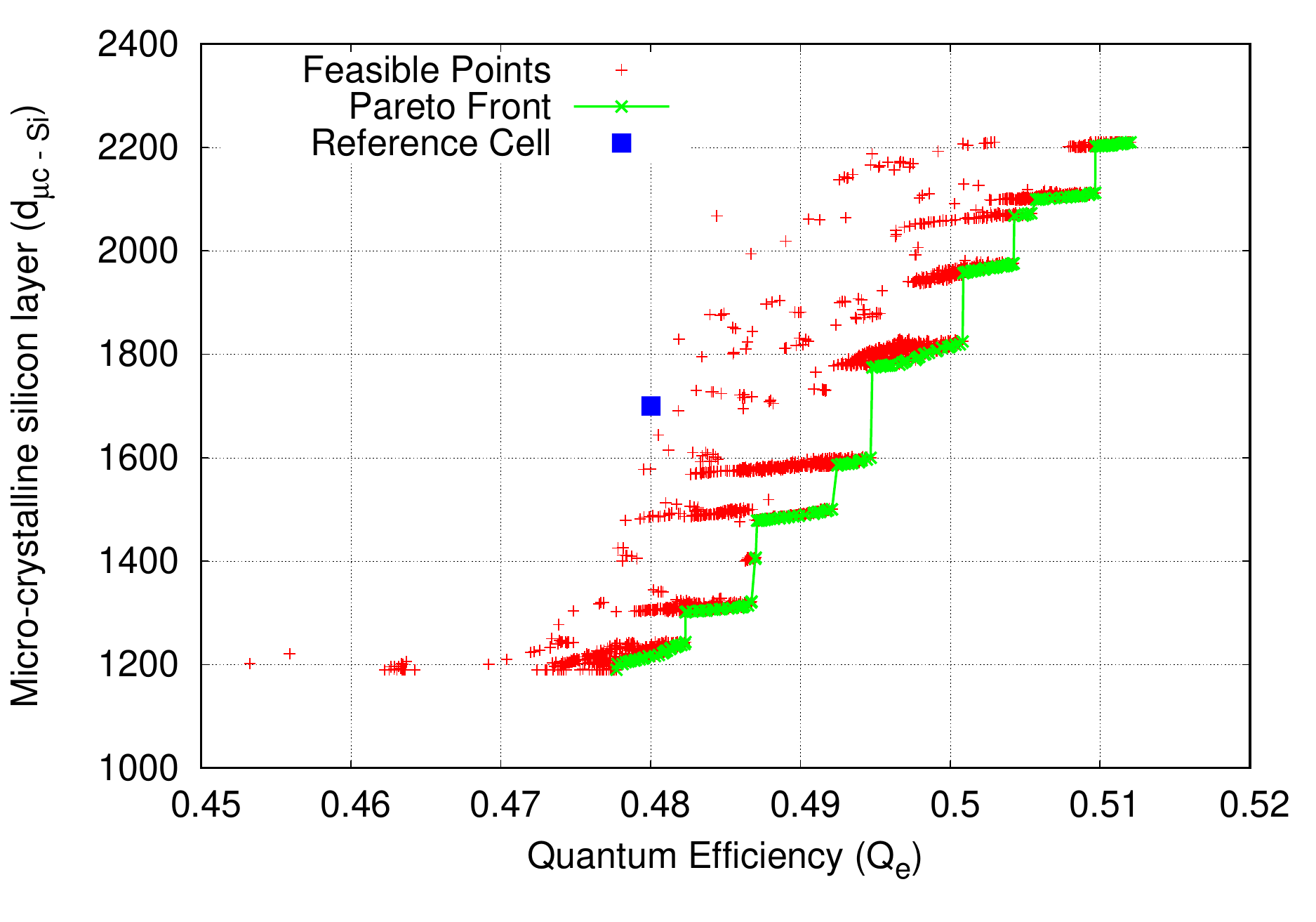}\label{fig:rAg_ntZnO}}
    	\subfigure[]{\includegraphics[width=0.32\columnwidth]{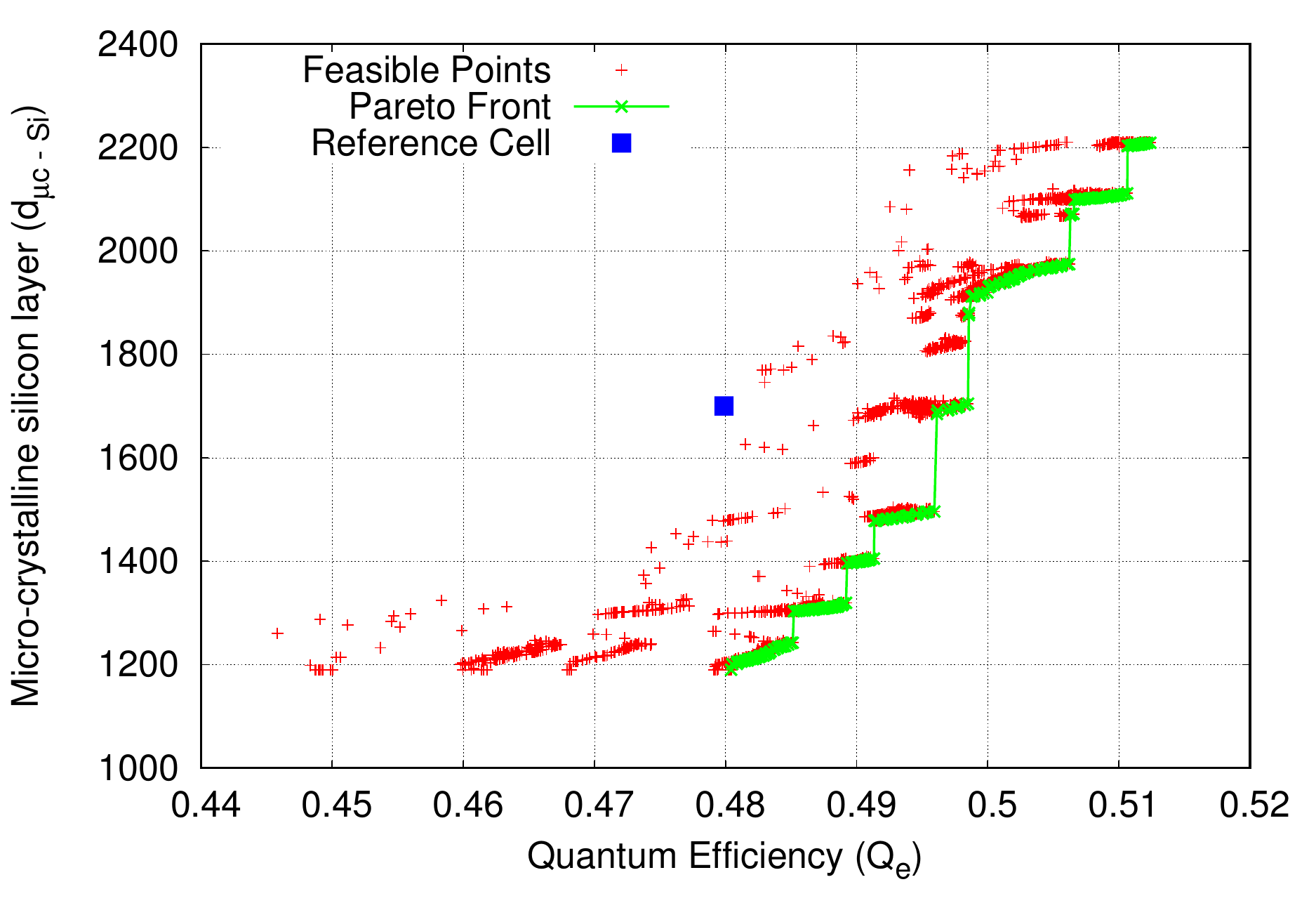}\label{fig:sAg_ntZnO}}
    	\subfigure[]{\includegraphics[width=0.32\columnwidth]{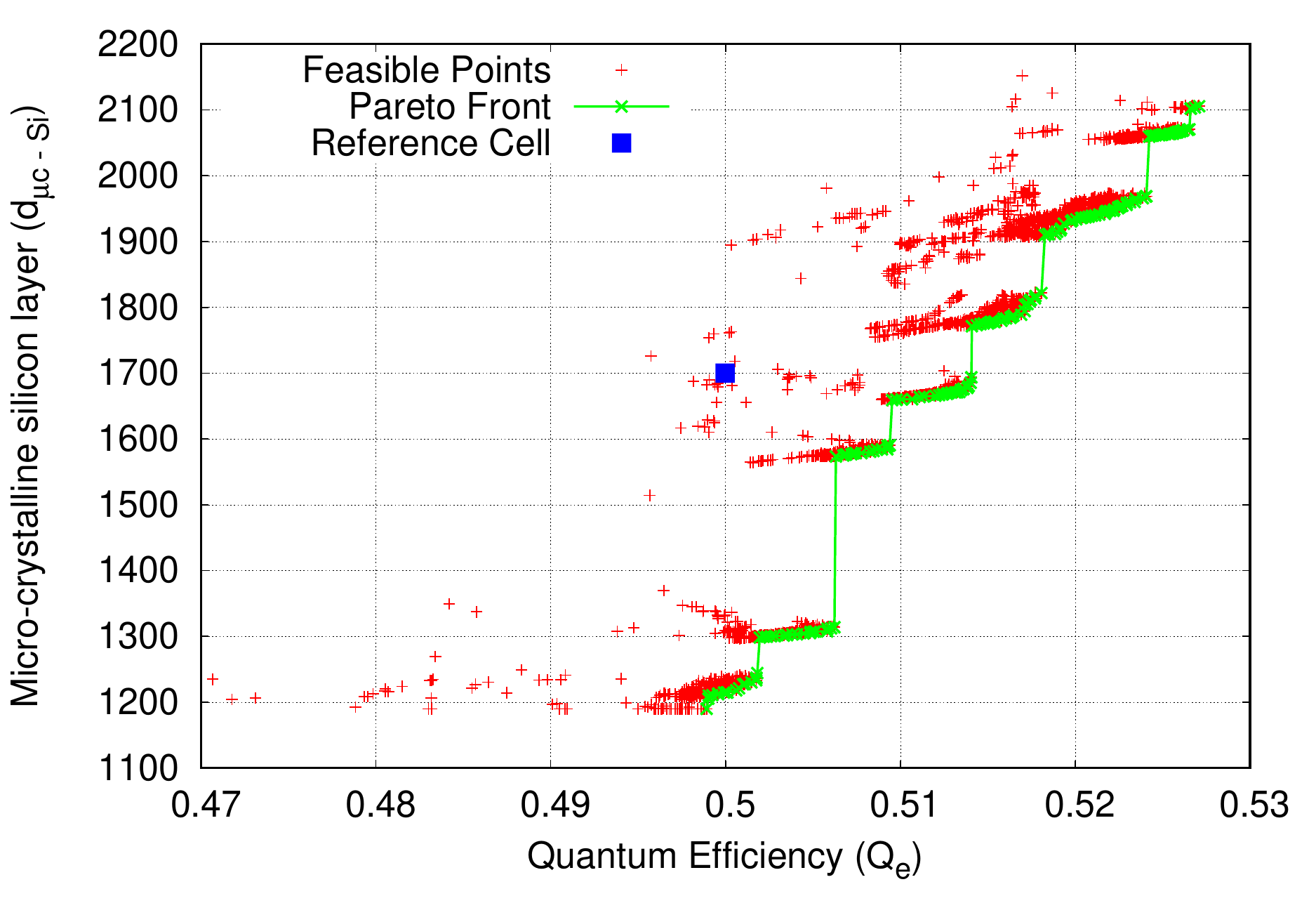}\label{fig:sAl_lZnO}}
    	\subfigure[]{\includegraphics[width=0.32\columnwidth]{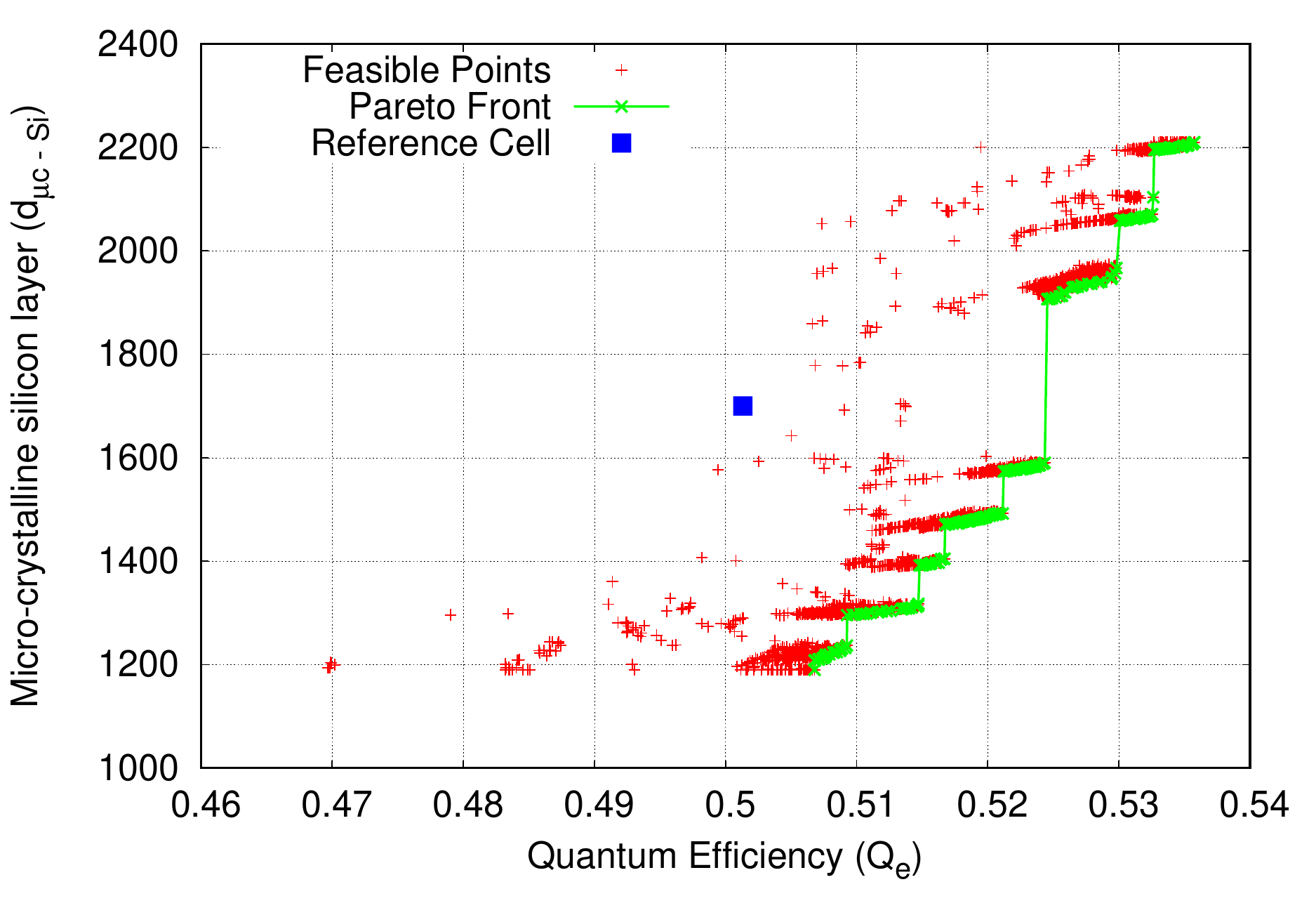}\label{fig:sAg_lZnO}}
    	\subfigure[]{\includegraphics[width=0.32\columnwidth]{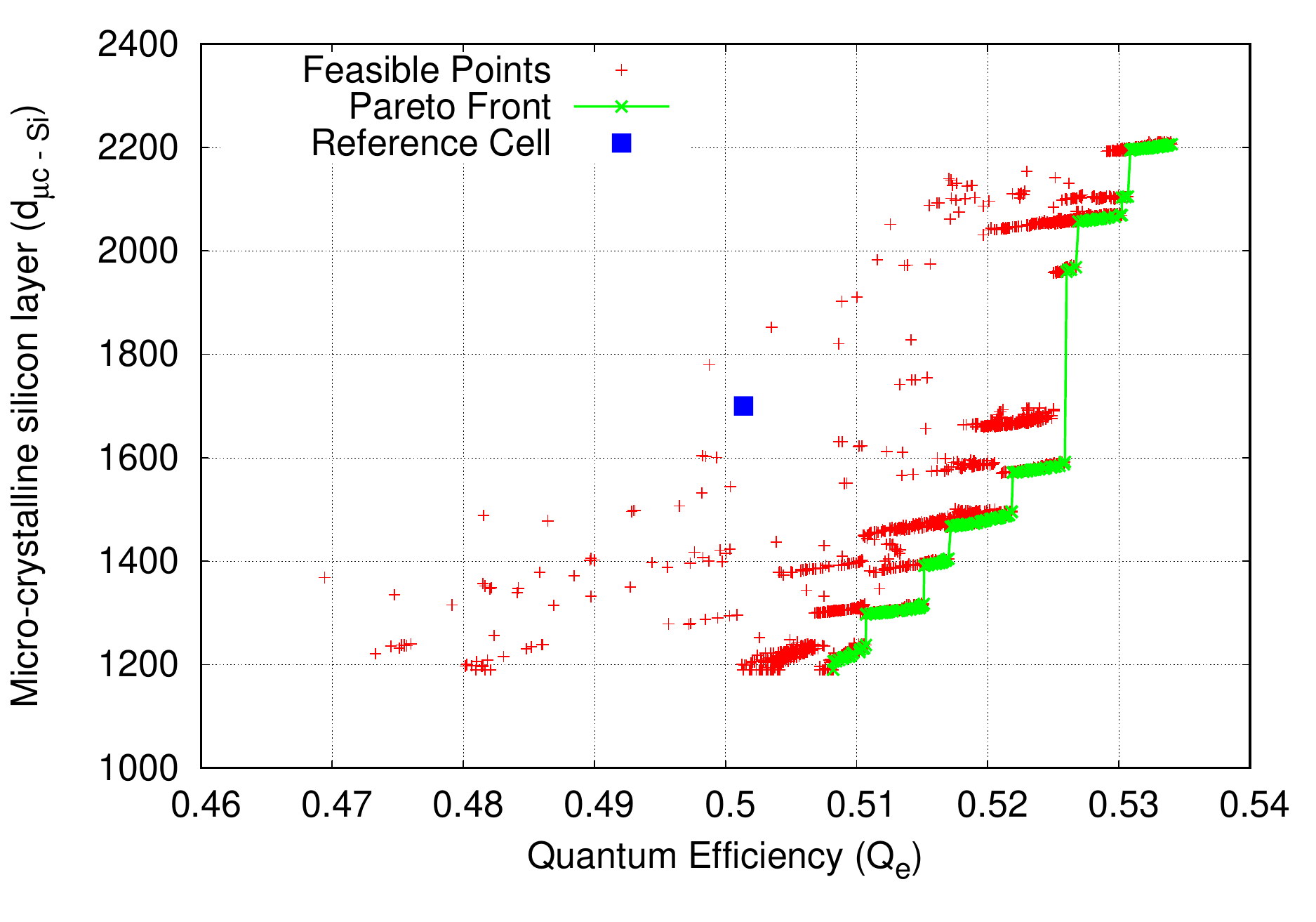}\label{fig:rAg_lZnO}}
    	\subfigure[]{\includegraphics[width=0.32\columnwidth]{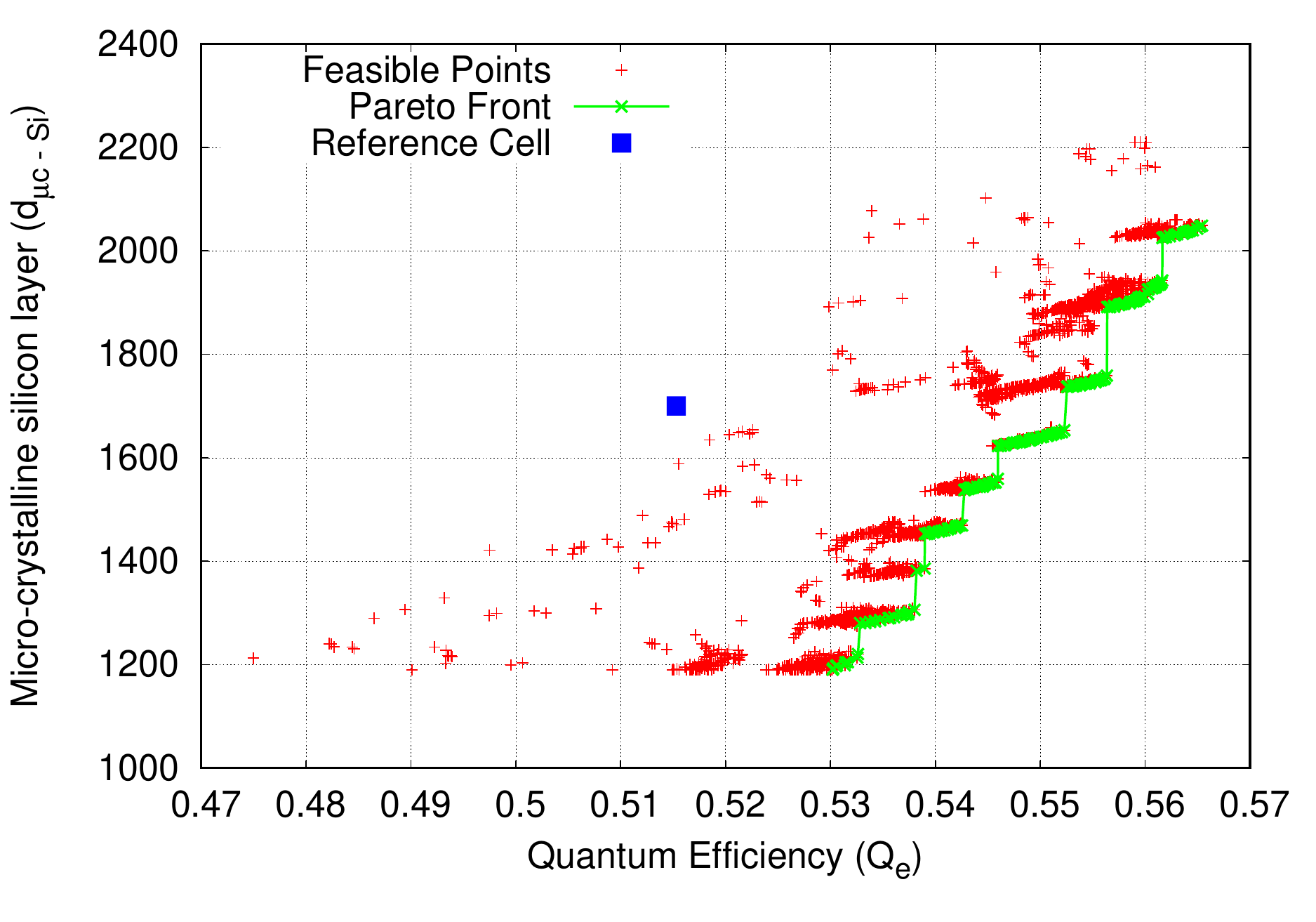} \label{fig:sAl_nSnO2}}
    	\subfigure[]{\includegraphics[width=0.32\columnwidth]{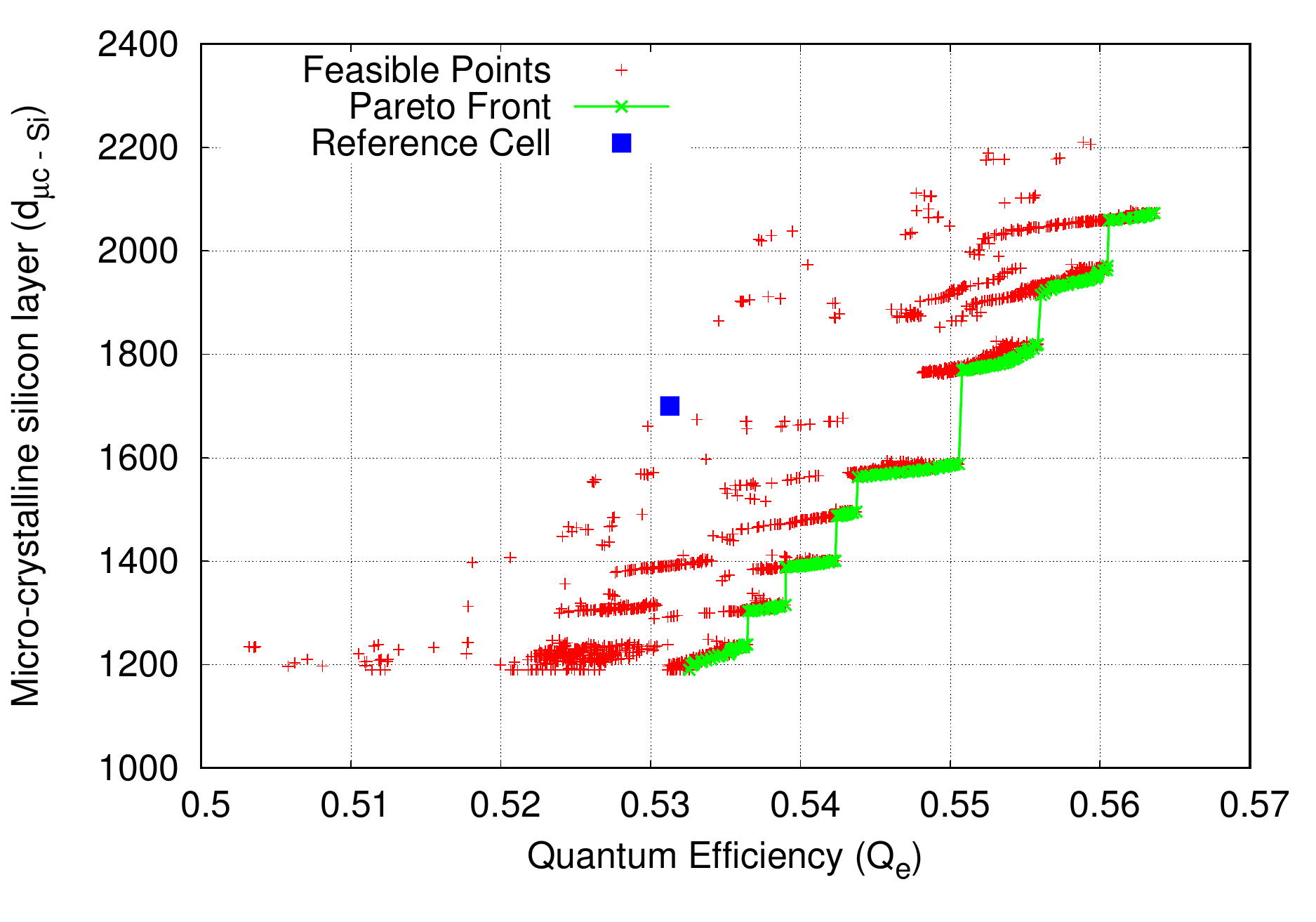}\label{fig:sAl_nZnO}}
    	\subfigure[]{\includegraphics[width=0.32\columnwidth]{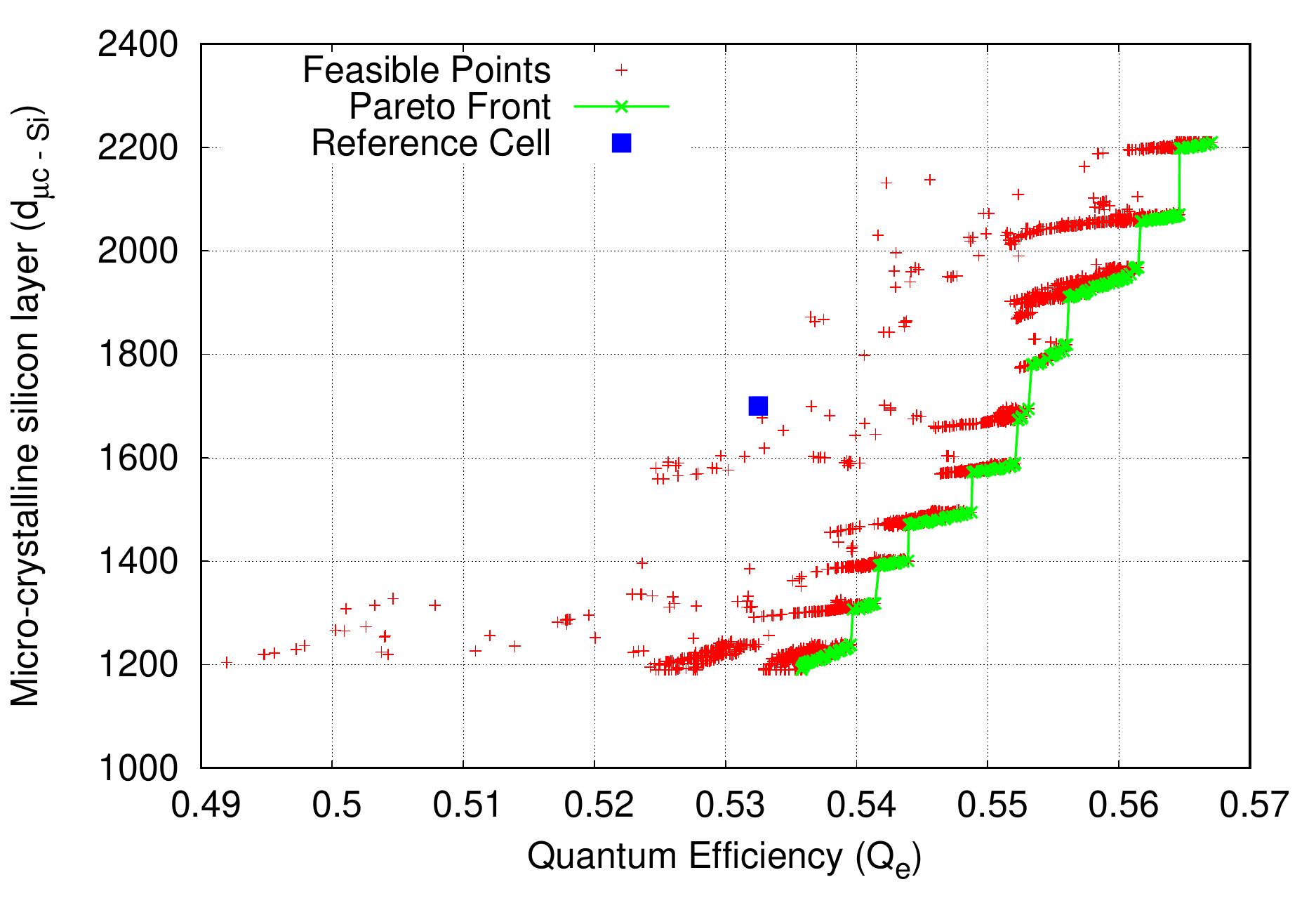}\label{fig:sAg_nZnO}}
    	
    	\subfigure[]{\includegraphics[width=0.32\columnwidth]{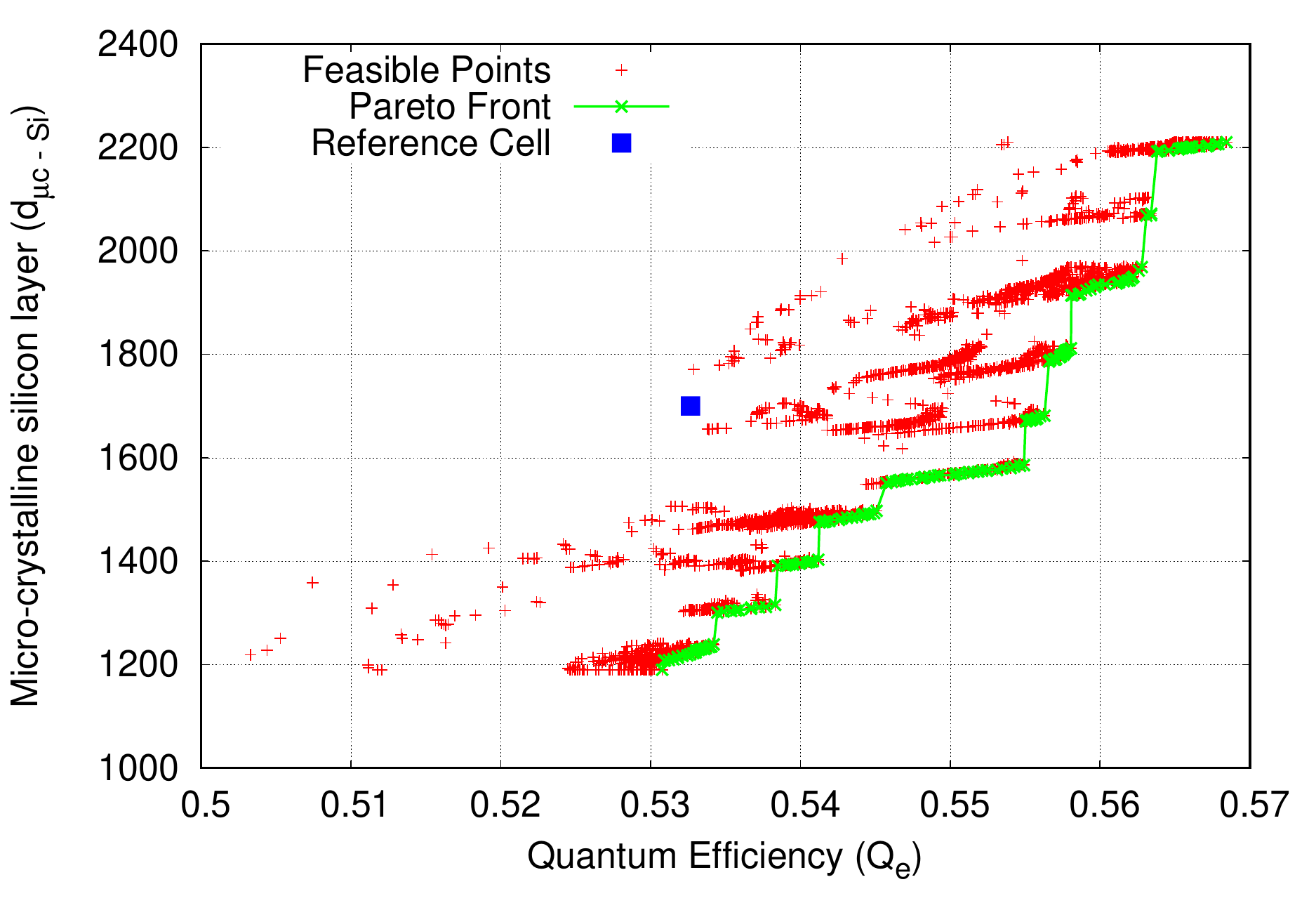}\label{fig:rAg_nZnO}}
    	\subfigure[]{\includegraphics[width=0.32\columnwidth]{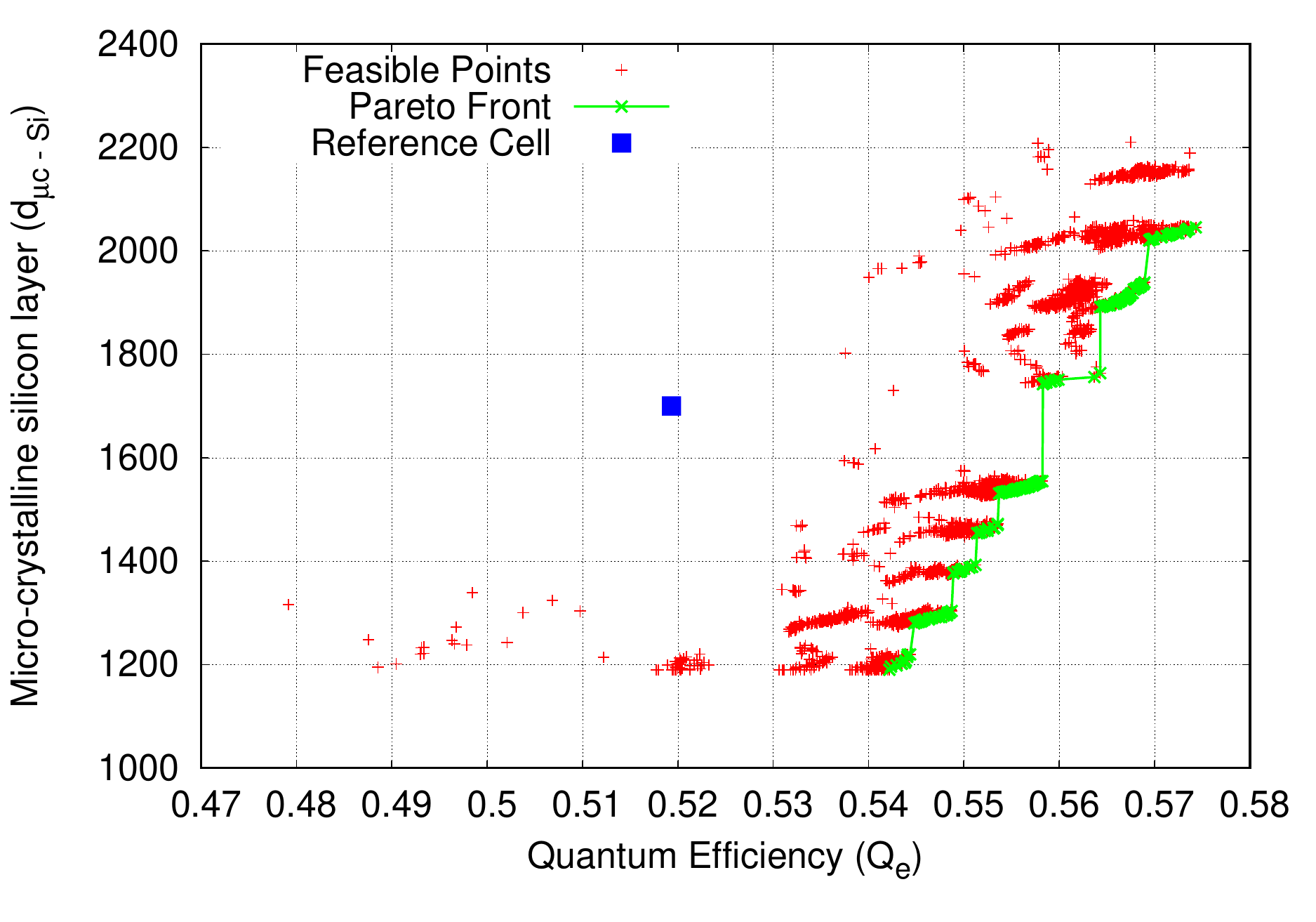}\label{fig:sAg_nSnO2}}
    	\subfigure[]{\includegraphics[width=0.32\columnwidth]{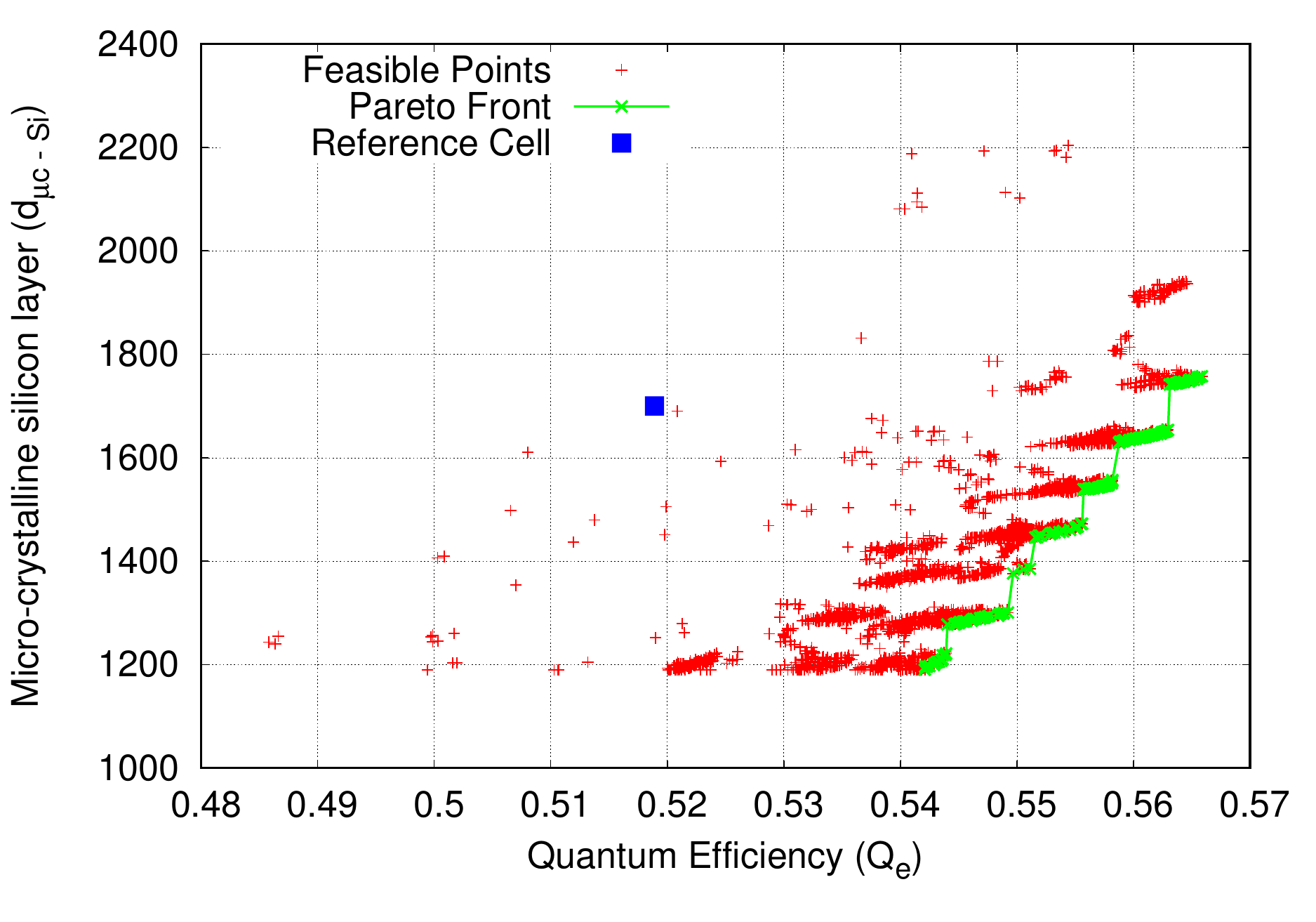}\label{fig:rAg_nSnO2}}
    	
    	\caption{NSGA-II optimization of simulation. (a)--(c) cluster of \textit{not-doped} ZnO cell structure designs for smooth back reflector Al,  rough back reflector Ag, smooth back reflector Ag. (d)--(f) cluster of \textit{lowly-doped} ZnO cell structure designs for smooth back reflector Al, smooth back reflector Ag, and rough back reflector Ag. (g)--(f) cluster of \textit{normally-doped} ZnO and SnO$_2$ doped cell structure designs for smooth back reflector Al (SnO$_2$), smooth back reflector Al (ZnO), smooth back reflector Ag (ZnO), rough back reflector Ag (ZnO),  smooth back reflector Ag (SnO$_2$), and rough back reflector Ag (SnO).}
    	\label{fig:other_clusters}
    \end{figure}

    NSGA-II finds optimal parameters for \textit{not-doped} designs that maximized quantum efficiencies of these designs with respect to their respective reference cells (cf. Figs.~\ref{fig:sAl_ntZnO}, \ref{fig:rAg_ntZnO}, and \ref{fig:sAg_ntZnO}). 
    Table~\ref{tab:simulation_sigma} summarizes parameter values of candidate solutions obtained for \textit{not-doped} cell designs: sAg-ntZnO, rAg-ntZnO, and sAl-ntZnO.
    Moreover, Table~\ref{tab:simulation_sigma} provides a detailed optimized parameter (layers interface roughness $ \sigma_i, i=2,\ldots,11 $) values and corresponding quantum efficiency, $Q_e$ values and microcrystalline silicon layer thickness, $d_{\mu c-Si}$ values of \textit{three candidate solution} points for each design.

    These three candidate solutions refer to 
    (i) the highest quantum efficiency $Q_e(\boldsymbol\sigma)^*$;
    (ii) the thinnest microcrystalline silicon layer $d_{\mu c-Si}(\boldsymbol\sigma)^*$; and
    (iii) a point $U(\boldsymbol\sigma)^*$ with the closest fitness to the \textit{Utopian point}, that is, the theoretical point having the best obtained optimal values for both objectives simultaneously.
    The Utopian point is not practically feasible. However, the presented $U(\boldsymbol\sigma)^*$ gives a good trade-off between the objectives in our simulations with respect to respective reference cell.

    \paragraph{Low cost-intensive and low quantum efficient cluster.}
    The next cluster in Fig.~\ref{fig:simulation_all_results} is \textit{lowly-doped} cell designs (sAl-lZnO, sAg-lZnO, and rAg-lZnO) whose quantum efficiency $Q_e$ varies between $0.5$ and $0.54$. 
    Their detailed Pareto-fronts are shown in Figs.~\ref{fig:sAl_lZnO}, 
    \ref{fig:sAg_lZnO}, and \ref{fig:rAg_lZnO}. 
    Table~\ref{tab:simulation_sigma} summarizes their parameter values of candidate solutions.

    \paragraph{High cost-intensive and high quantum efficient cluster.}
    In the order of the performances, the design with rough BR rAg-lZnO is better than smooth doping designs, which indicates that low doping rough BR cell design provides better light-harvesting capacity than smooth BR cell design. 
    This fact is evident from \textit{normally-doped} ZnO design and SnO$_2$ doped design cluster (cf. Fig.~\ref{fig:simulation_all_results}). 
    The detailed Pareto-fronts of \textit{normally-doped} cluster designs are shown in Figs.~\ref{fig:sAl_nSnO2}, \ref{fig:sAl_nZnO}, \ref{fig:sAg_nZnO}, \ref{fig:rAg_nZnO}, \ref{fig:sAg_nSnO2}, and \ref{fig:rAg_nSnO2} respectively. 
    Their respective parameter values of three candidate solutions are shown in Table~\ref{tab:simulation_sigma}. 
    
    After analysing \textit{lowly-doped} and \textit{normally-doped} clusters, it is clearly observed that the rough BR Ag among ZnO based TCO doping is a better choice than a smooth BR. 
    Similarly, rough BR Ag gives better light-harvesting capacity than a smooth BR.
    In-fact, SnO$_2$ based TCO layer shows a higher quantum efficiency than ZnO based TCO layer in this cluster. 
    The max quantum efficiency of rAg-SnO$_2$ is $0.57$, which is as good as the reference cell outlined in Table~\ref{tab:refernce_cell}.

\begin{figure}\centering
        \includegraphics[width=0.98\columnwidth]{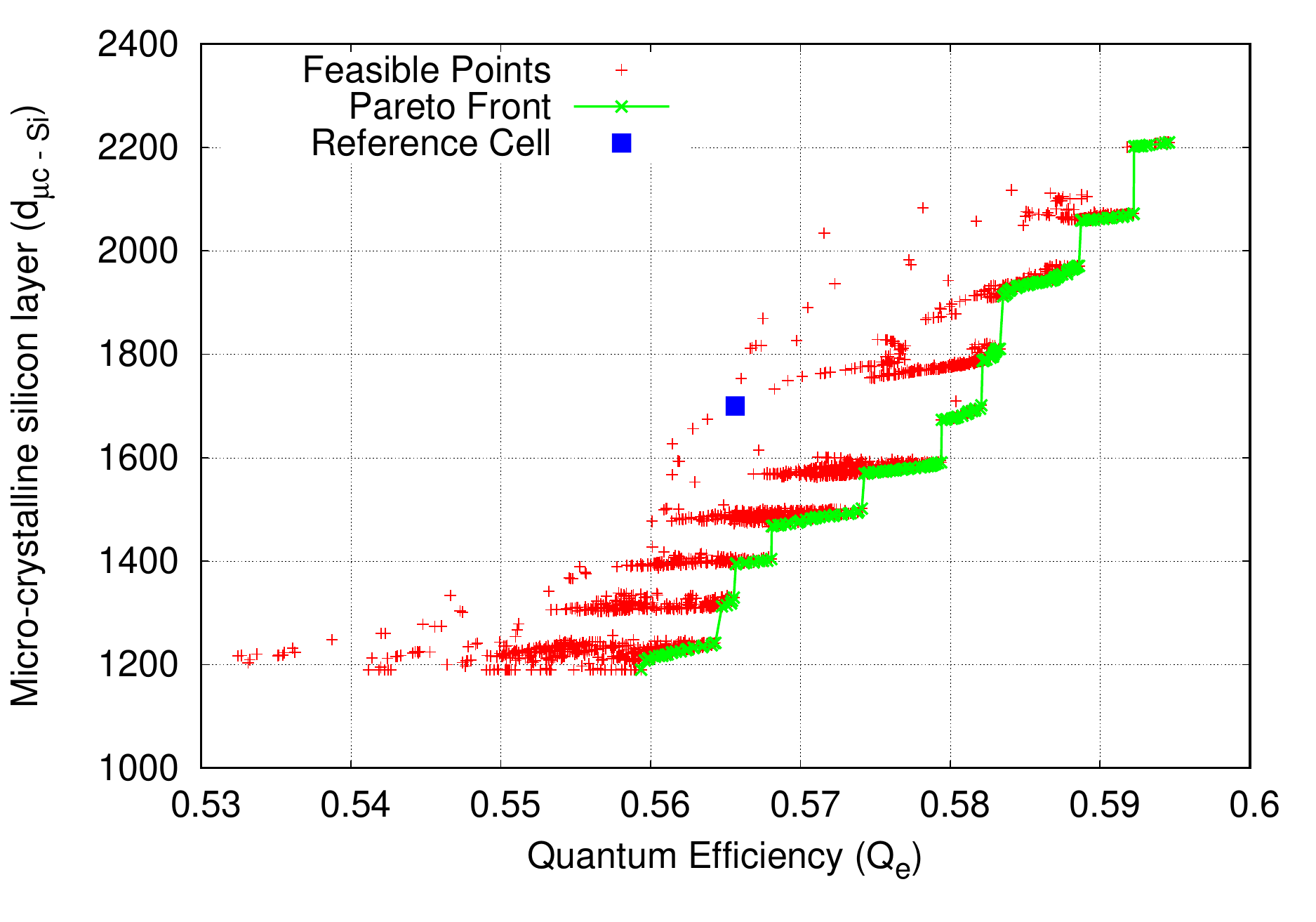}
        \caption{\label{fig:select_sAl_oZnO}
        	NSGA-II optimization of smooth back reflector Al plus optimally doped ZnO cell design. The points $Q_e(\boldsymbol\sigma)^*$ is (0.6, 2210), $d_{\mu c-Si}(\boldsymbol\sigma)^*$ is (0.56, 1190) and $U(\boldsymbol\sigma)^*$ is (0.58, 1586.92).}
    \end{figure}

\begin{figure}\centering
		\includegraphics[width=0.98\columnwidth]{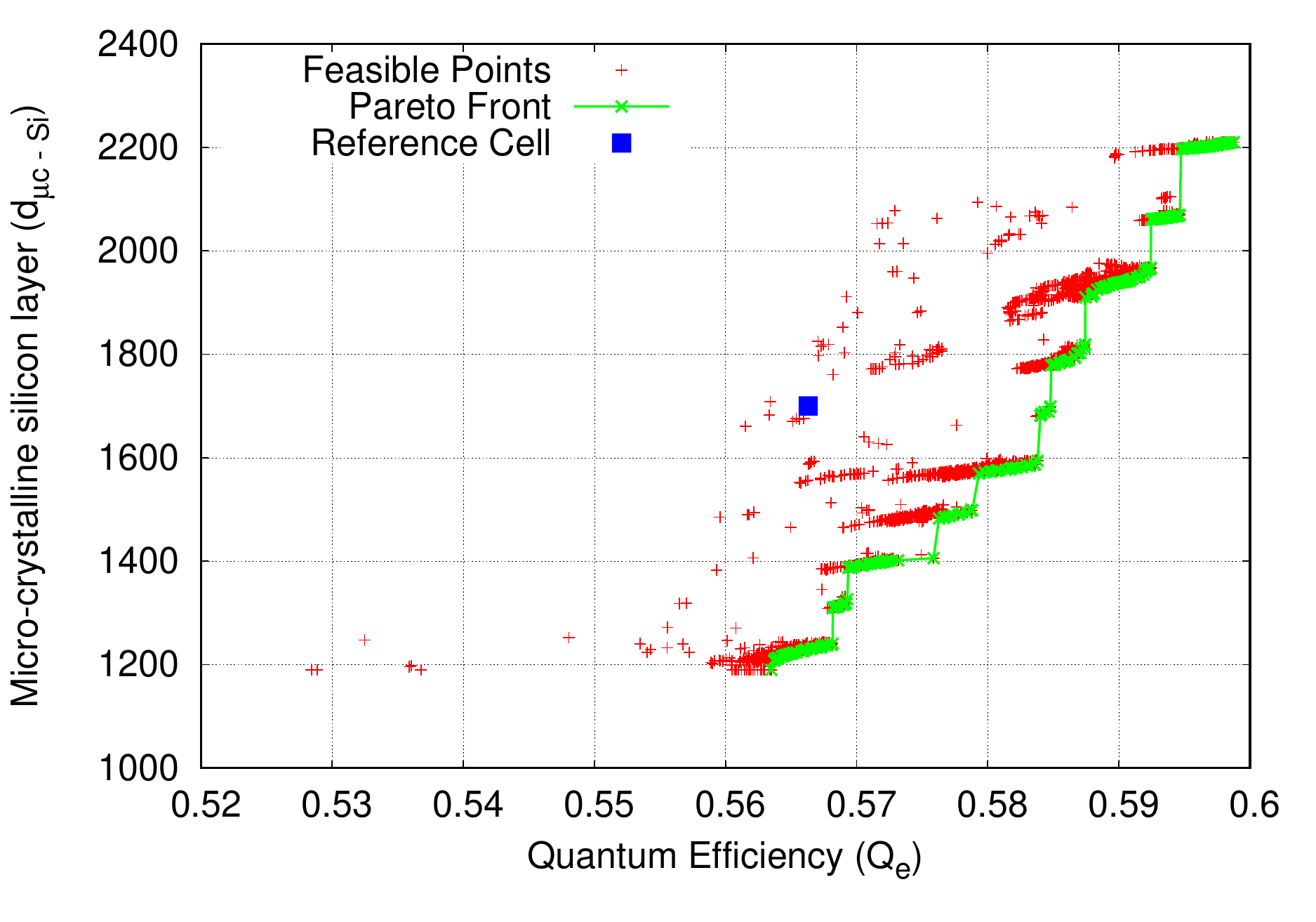}
		\caption{\label{fig:select_sAg_oZnO}
			NSGA-II optimization of smooth back reflector Ag plus optimally doped ZnO cell design. The points $Q_e(\boldsymbol\sigma)^*$ is (0.59, 2210), $d_{\mu c-Si}(\boldsymbol\sigma)^*$ is (0.56, 1190) and $U(\boldsymbol\sigma)^*$ is (0.58, 1590.91).}
	\end{figure}

\begin{figure}\centering
        \includegraphics[width=0.98\columnwidth]{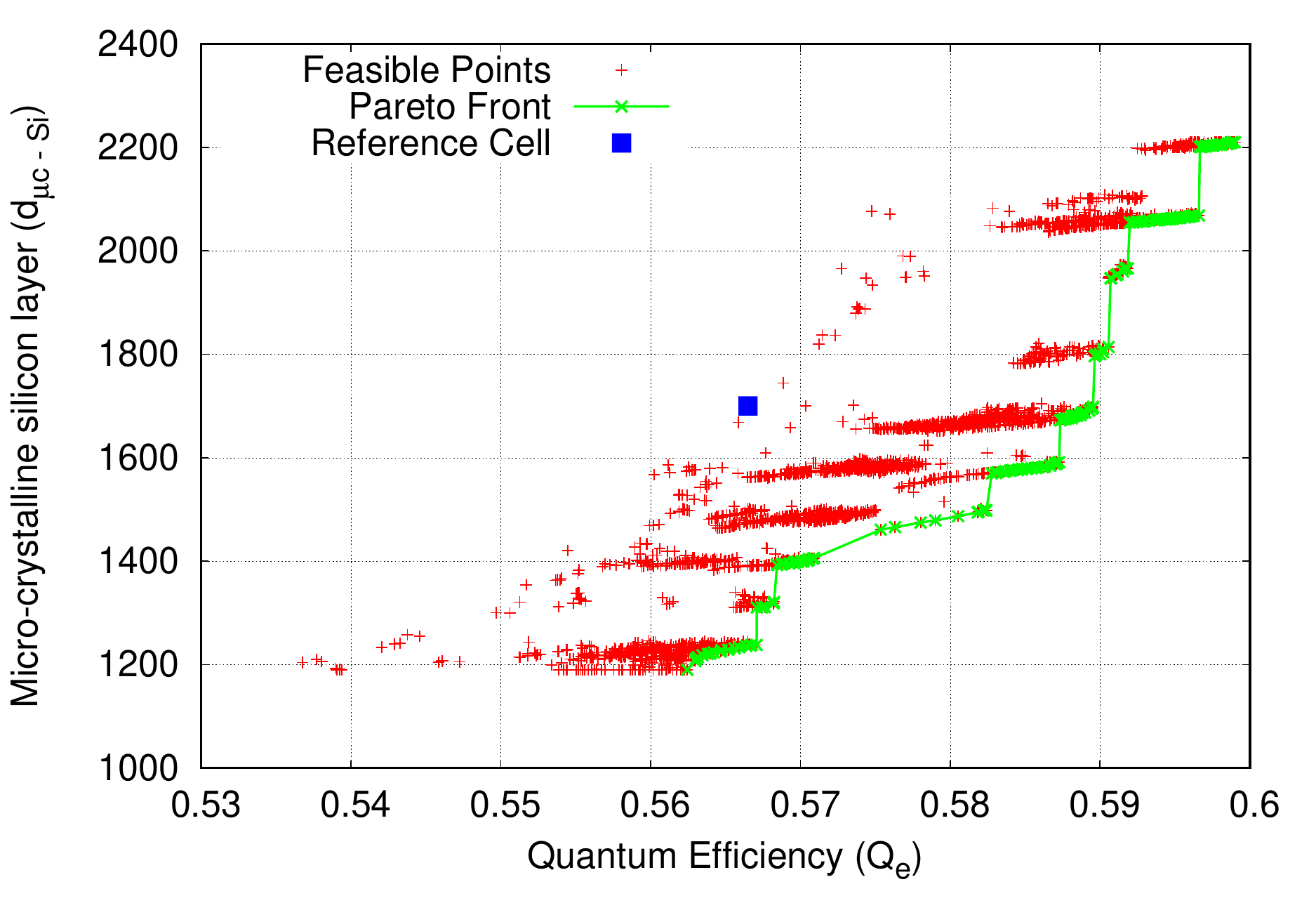}
        \caption{\label{fig:select_rAg_oZnO}
        	NSGA-II optimization of rough back reflector Ag plus optimally doped ZnO cell design. The points $Q_e(\boldsymbol\sigma)^*$ is (0.6, 2210), $d_{\mu c-Si}(\boldsymbol\sigma)^*$ is (0.56, 1190) and $U(\boldsymbol\sigma)^*$ is (0.59, 1587.99).}
    \end{figure}

    \paragraph{Highest cost-intensive and highest quantum efficient cluster.}
    The rightmost cluster in Fig.~\ref{fig:simulation_all_results} is the cluster of \textit{optimally doped} cell designs: ``smooth BR Al plus optimally doped ZnO (sAl-oZnO),'' ``smooth BR Ag plus optimally doped ZnO (sAg-oZnO),'' and ``rough BR Ag plus optimally doped ZnO (rAg-oZnO).'' 
    Figs.~\ref{fig:select_sAl_oZnO},  \ref{fig:select_sAg_oZnO}, and \ref{fig:select_rAg_oZnO}, respectively show detailed NSGA-II based MOO of the simulation and all feasible solutions on the Pareto-front of sAl-oZnO, sAg-oZnO and rAg-oZnO cell designs.

    The quantum efficiency of this cluster varies between $0.56$ (close to the reference cell) and $0.6$ ($\approx 5.36$\% more efficient than the reference cell and equal to baseline quantum efficiency $Q_e$). 
    Detailed parameter values and three candidate solutions of these designs are listed in Table~\ref{tab:simulation_sigma}. 
    We select these optimally doped cell designs (sAl-oZnO, sAg-oZnO, and rAg-oZnO) for the second stage optimization (Pareto-front's fine-tuning) using OptIA-II. 
    
	\begin{table}
		\scriptsize
		\centering
		\renewcommand{\arraystretch}{1}
		\setlength{\tabcolsep}{3pt}
		\caption{\label{tab:simulation_sigma}
			\emph{NSGA-II} Multi-objective optimization (minimization of Micro-crystal silicon layer $d_{\mu c - Si}$ and maximization of Quantum efficiency $Q_e$) of the optical model for tandem-thin-film silicon solar cell with the ``full parameterized'' model for all 15 solar cell simulation variations. 
			For each setting, the two extreme points $Q_e(\boldsymbol\sigma)^*$ and $d_{\mu c-Si}(\boldsymbol\sigma)^*$ and the point $U(\boldsymbol\sigma)^*$ closest to the utopian point are reported. 
			Highlighted bold $Q_e$ values are three select candidate designs whose Pareto-front was passed onto the second stage for the optimization.}
		\begin{tabular}{ccccccccccccc}
			\toprule
			Simulation & $\sigma_2$ & $\sigma_3$ & $\sigma_4$ & $\sigma_5$ & $\sigma_6$ & $\sigma_7$ & $\sigma_8$ & $\sigma_9$ & $\sigma_{10}$ & $\sigma_{11}$ & $d_{\mu c-Si}$ & $Q_e$\\
			\midrule
			\multirow{3}*{sAl-ntZnO} 
            & 143.57 & 1.52 & 59.73 & 0.26 & 1.13 & 2.57 & 228.51 & 0.81 & 3.92 & 7.35 & 2210.00 & 0.51\\
            & 146.91 & 0.97 & 58.20 & 3.12 & 2.06 & 3.07 & 98.00 & 0.76 & 3.85 & 30.00 & 1501.87 & 0.49\\
            & 147.34 & 1.07 & 60.00 & 3.42 & 2.49 & 3.62 & 299.02 & 3.83 & 4.00 & 9.28 & 1190.00 & 0.47\\
            & & & & & & & & & & & & \\
            
            \multirow{3}*{rAg-ntZnO} 
            & 148.95 & 1.35 & 59.07 & 0.20 & 1.37 & 2.65 & 241.37 & 1.24 & 3.99 & 4.51 & 2210.00 & 0.51\\
            & 176.18 & 3.99 & 53.07 & 0.61 & 2.16 & 3.93 & 299.62 & 4.00 & 1.68 & 18.50 & 1600.04 & 0.49\\
            & 140.97 & 2.11 & 59.82 & 1.54 & 2.52 & 1.60 & 109.31 & 1.59 & 3.85 & 29.54 & 1190.00 & 0.48\\
            & & & & & & & & & & & & \\
            
            \multirow{3}*{sAg-ntZnO} 
            & 149.70 & 1.26 & 59.88 & 0.61 & 1.07 & 2.80 & 221.45 & 1.12 & 4.00 & 5.96 & 2209.17 & 0.51\\
            & 170.43 & 3.87 & 56.98 & 0.55 & 0.27 & 2.46 & 301.44 & 2.27 & 3.98 & 29.67 & 1495.93 & 0.50\\
            & 179.78 & 0.20 & 58.98 & 3.72 & 2.51 & 1.84 & 233.94 & 4.00 & 3.98 & 18.34 & 1190.00 & 0.48\\
            & & & & & & & & & & & & \\
            
            \multirow{3}*{sAl-lZnO} 
            & 163.20 & 1.94 & 54.36 & 0.64 & 4.00 & 3.31 & 97.18 & 3.84 & 1.56 & 13.89 & 2105.49 & 0.53\\
            & 166.56 & 3.12 & 49.41 & 0.87 & 3.02 & 0.42 & 201.36 & 3.70 & 1.83 & 20.79 & 1677.59 & 0.51\\
            & 155.36 & 3.79 & 51.13 & 2.33 & 3.03 & 3.84 & 313.33 & 1.07 & 4.00 & 29.27 & 1190.00 & 0.50\\
            & & & & & & & & & & & & \\
            
            \multirow{3}*{sAg-lZnO} 
            & 178.73 & 1.78 & 53.60 & 2.90 & 1.03 & 1.64 & 250.36 & 2.02 & 2.00 & 19.61 & 2210.00 & 0.54\\
            & 179.90 & 1.03 & 59.96 & 2.56 & 1.12 & 3.99 & 118.19 & 2.84 & 3.39 & 10.19 & 1590.20 & 0.52\\
            & 175.71 & 1.07 & 59.54 & 2.71 & 0.96 & 3.77 & 141.00 & 2.85 & 3.74 & 12.75 & 1190.00 & 0.51\\
            & & & & & & & & & & & & \\
            
            \multirow{3}*{rAg-lZnO} 
            & 168.72 & 1.15 & 54.17 & 1.00 & 0.23 & 0.53 & 117.43 & 0.86 & 1.61 & 24.84 & 2206.33 & 0.53\\
            & 180.00 & 1.18 & 57.30 & 0.45 & 3.89 & 3.05 & 121.30 & 2.12 & 3.94 & 29.95 & 1587.14 & 0.53\\
            & 177.03 & 1.26 & 57.80 & 0.57 & 3.82 & 2.31 & 112.01 & 2.55 & 3.92 & 27.28 & 1190.00 & 0.51\\
            & & & & & & & & & & & & \\

            \multirow{3}*{sAl-SnO2} 
            & 180.00 & 1.54 & 60.00 & 2.21 & 2.32 & 1.06 & 32.74 & 3.99 & 2.01 & 6.09 & 2048.73 & 0.57\\
            & 179.77 & 2.01 & 50.37 & 2.18 & 2.20 & 0.29 & 62.02 & 3.96 & 2.38 & 2.81 & 1653.14 & 0.55\\
            & 177.85 & 3.43 & 41.74 & 2.26 & 3.07 & 2.91 & 221.17 & 1.28 & 3.89 & 9.76 & 1190.00 & 0.53\\
            & & & & & & & & & & & & \\
            
            \multirow{3}*{sAl-nZnO} 
            & 179.68 & 0.63 & 59.99 & 3.03 & 2.18 & 3.74 & 306.67 & 3.23 & 3.23 & 3.04 & 2073.04 & 0.56\\
            & 179.86 & 1.30 & 55.95 & 3.50 & 3.42 & 2.42 & 70.84 & 4.00 & 1.50 & 29.77 & 1587.75 & 0.55\\
            & 179.95 & 1.46 & 59.95 & 3.65 & 3.73 & 2.34 & 60.03 & 4.00 & 1.22 & 22.20 & 1190.00 & 0.53\\
            & & & & & & & & & & & & \\
            
            \multirow{3}*{sAg-nZnO} 
            & 156.56 & 2.21 & 60.00 & 0.62 & 0.63 & 1.56 & 93.53 & 2.46 & 3.46 & 21.94 & 2210.00 & 0.57\\
            & 164.63 & 0.35 & 54.18 & 2.86 & 2.55 & 3.27 & 94.23 & 0.63 & 4.00 & 28.07 & 1586.12 & 0.55\\
            & 180.00 & 2.62 & 52.79 & 2.86 & 0.30 & 0.55 & 167.91 & 3.23 & 3.63 & 27.46 & 1190.00 & 0.54\\
            & & & & & & & & & & & & \\

			\multirow{3}*{rAg-nZnO} 
            & 168.86 & 1.66 & 59.66 & 0.82 & 3.86 & 3.45 & 140.12 & 3.43 & 2.57 & 9.84 & 2210.00 & 0.57\\
            & 178.50 & 3.37 & 53.61 & 1.55 & 2.80 & 1.36 & 323.05 & 2.48 & 3.99 & 30.00 & 1585.69 & 0.55\\
            & 153.24 & 2.67 & 55.05 & 0.20 & 2.90 & 2.21 & 333.86 & 0.82 & 4.00 & 9.29 & 1190.00 & 0.53\\
            & & & & & & & & & & & & \\
            
            \multirow{3}*{sAg-SnO2} 
            & 180.00 & 1.89 & 49.77 & 3.05 & 1.43 & 3.70 & 70.69 & 3.93 & 3.93 & 27.58 & 2045.57 & 0.57\\
            & 180.00 & 2.58 & 54.09 & 3.22 & 3.69 & 1.44 & 315.97 & 0.71 & 1.55 & 28.30 & 1554.78 & 0.56\\
            & 171.10 & 3.20 & 60.00 & 3.76 & 4.00 & 1.06 & 286.61 & 1.37 & 1.84 & 28.37 & 1190.00 & 0.54\\
            & & & & & & & & & & & & \\
            
            \multirow{3}*{rAg-SnO2} 
            & 180.00 & 3.47 & 60.00 & 0.39 & 2.99 & 3.17 & 132.56 & 0.38 & 3.34 & 7.77 & 1757.36 & 0.57\\
            & 179.93 & 0.73 & 54.96 & 3.29 & 1.25 & 1.78 & 30.78 & 3.93 & 3.18 & 12.06 & 1472.39 & 0.56\\
            & 171.10 & 2.83 & 56.54 & 2.80 & 2.57 & 4.00 & 194.27 & 0.91 & 3.70 & 24.37 & 1190.00 & 0.54\\
            & & & & & & & & & & & & \\
            
            \multirow{3}*{sAl-oZnO} 
            & 156.25 & 1.43 & 59.48 & 0.27 & 1.21 & 2.69 & 228.92 & 0.83 & 3.28 & 6.75 & 2210.00 & \textbf{0.59}\\
            & 179.24 & 1.29 & 50.42 & 3.45 & 0.79 & 2.31 & 297.25 & 3.78 & 4.00 & 8.29 & 1590.91 & 0.58\\
            & 148.89 & 3.00 & 53.89 & 3.23 & 1.17 & 3.96 & 126.97 & 1.89 & 4.00 & 12.81 & 1190.00 & 0.56\\
            & & & & & & & & & & & & \\
            
            \multirow{3}*{sAg-oZnO} 
			& 180.00 & 1.72 & 60.00 & 0.66 & 1.48 & 2.46 & 39.03 & 1.25 & 0.75 & 11.86 & 2210.00 & \textbf{0.60}\\
			& 168.33 & 0.49 & 59.87 & 0.35 & 4.00 & 3.66 & 93.60 & 0.33 & 2.42 & 28.01 & 1586.92 & 0.58\\
			& 161.10 & 3.41 & 56.46 & 2.20 & 1.71 & 3.22 & 255.27 & 1.17 & 3.51 & 12.91 & 1190.00 & 0.56\\
			& & & & & & & & & & & & \\
            
            \multirow{3}*{rAg-oZnO} 
            & 180.00 & 1.38 & 56.95 & 0.38 & 0.88 & 3.19 & 211.80 & 1.26 & 2.96 & 6.44 & 2210.00 & \textbf{0.60}\\
            & 180.00 & 1.03 & 59.68 & 1.23 & 3.91 & 3.11 & 41.17 & 1.28 & 3.66 & 30.00 & 1587.99 & 0.59\\
            & 158.65 & 3.27 & 52.08 & 2.67 & 1.86 & 0.89 & 326.33 & 1.19 & 3.84 & 28.74 & 1190.00 & 0.56\\
			\bottomrule
		\end{tabular}
	\end{table}

	\subsection{Stage 2: Multi-objective optimization using OptIA-II}
\begin{figure}
        \centering
        \includegraphics[width=0.98\columnwidth]{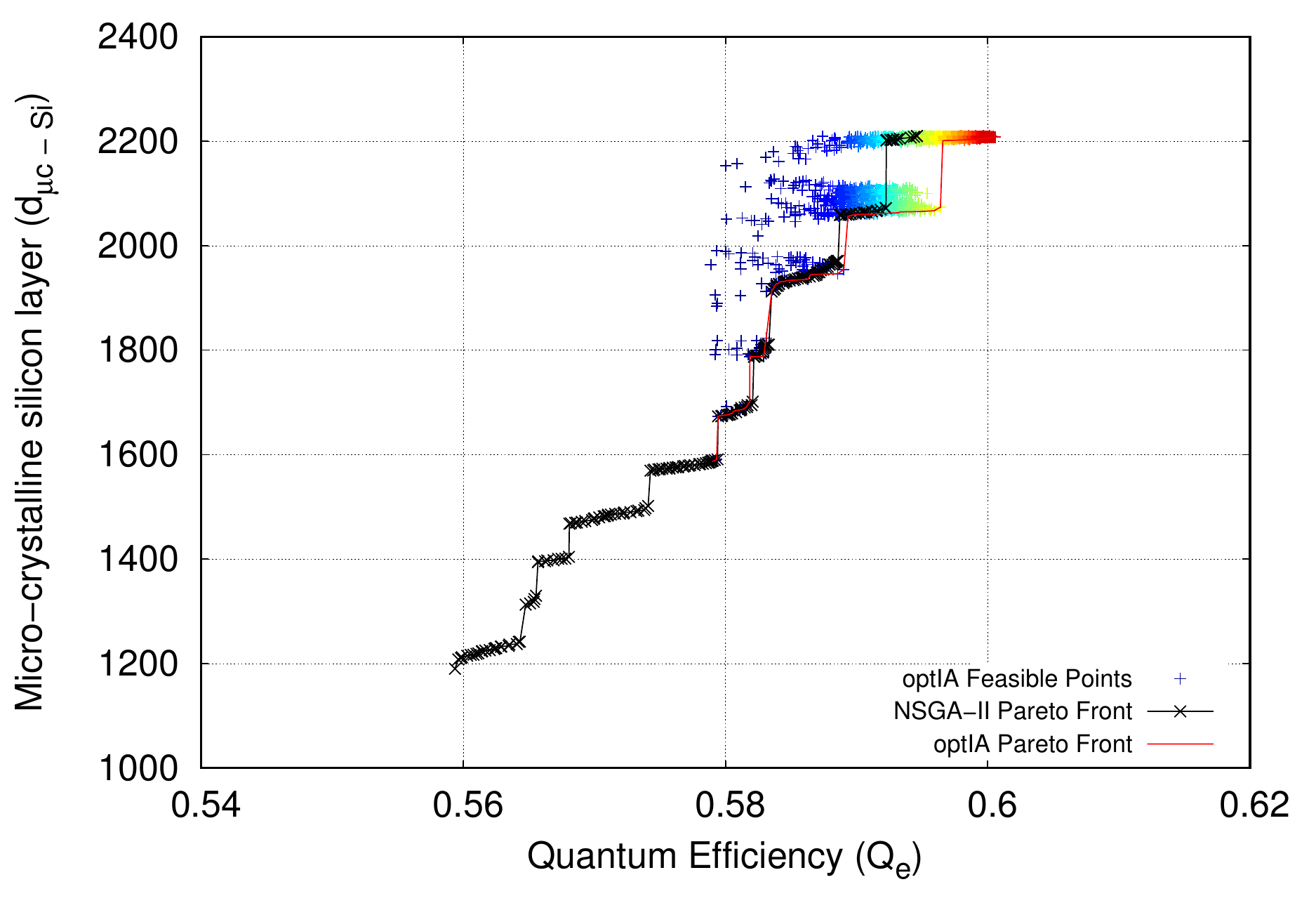}
        \caption{Second stage \emph{OptiA-II} optimization for Smooth back reflector Al, Optimally Doped ZnO. Pareto-front of \textit{OptIA-II} optimization is indicated in red.}
        \label{fig:select_sAl_oZnO_OPT}
    \end{figure}

\begin{figure}
		\centering
		\includegraphics[width=0.98\columnwidth]{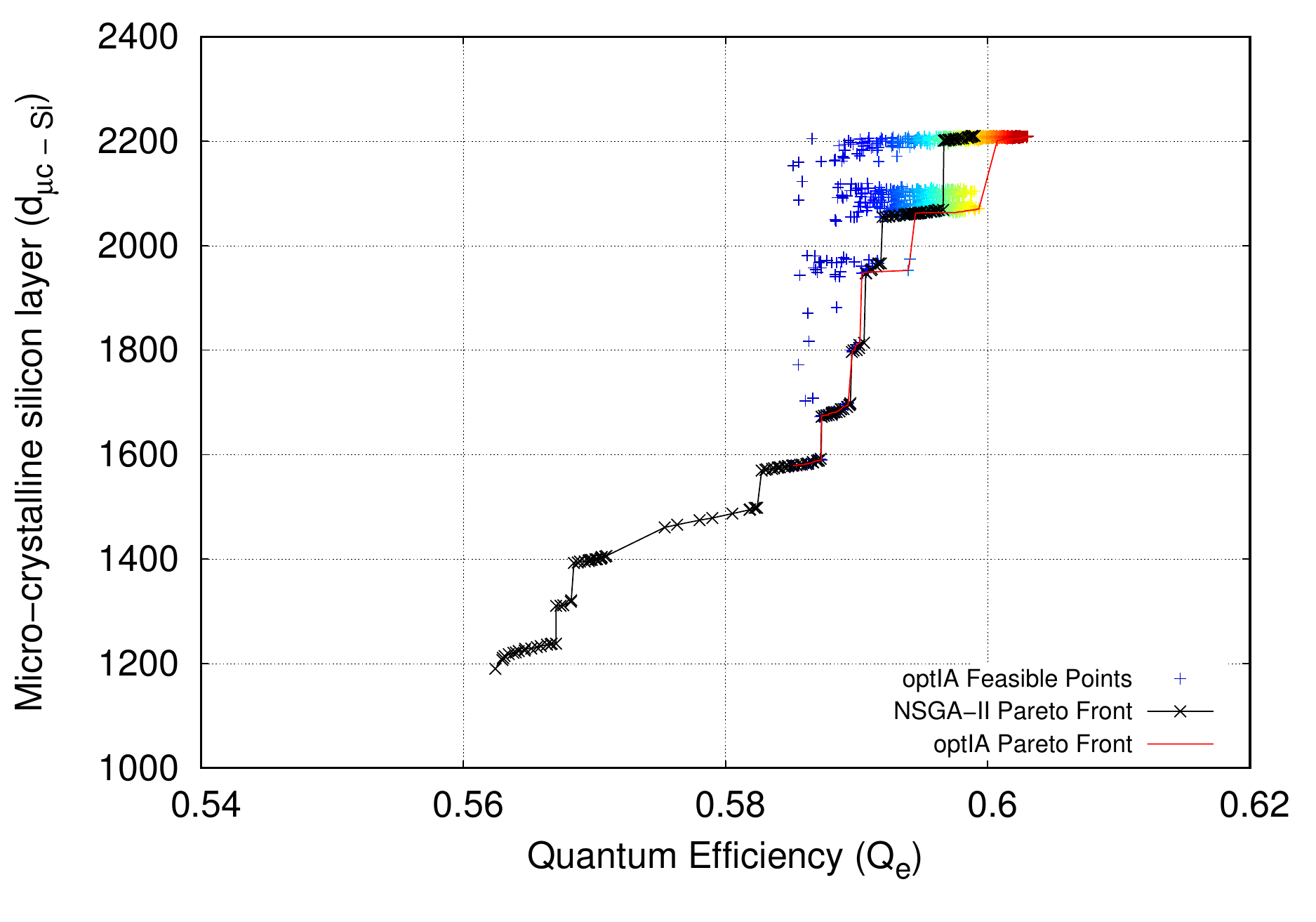}
		\caption{Second stage \emph{OptiA-II} optimization for Rough back reflector Ag, Optimally Doped ZnO. Pareto-front of \textit{OptIA-II} optimization is indicated in red.}
        \label{fig:select_sAg_oZnO_OPT}
	\end{figure}

\begin{figure}
		\centering
		\includegraphics[width=0.98\columnwidth]{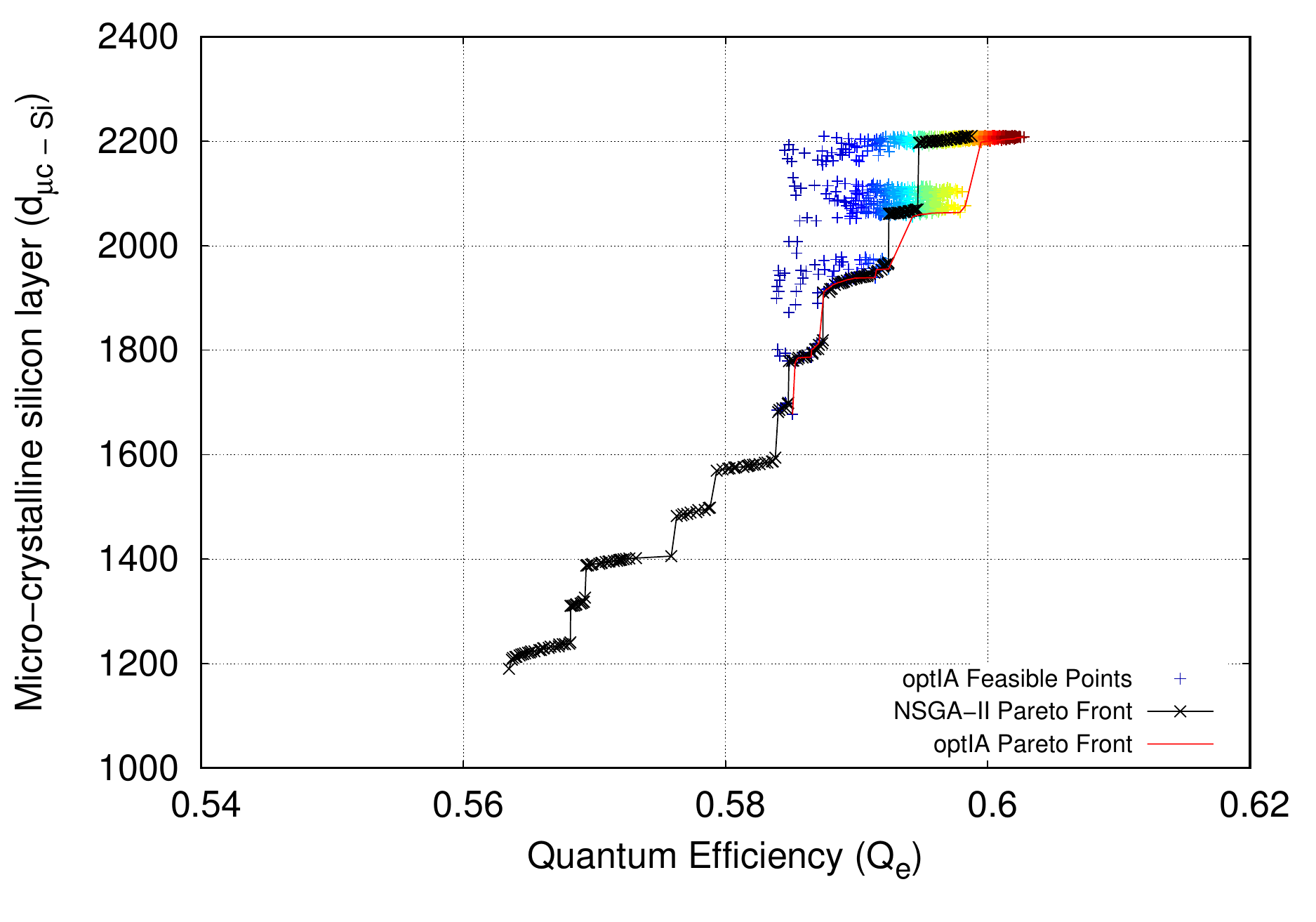}
		\caption{Second stage \emph{OptIA-II} optimization for Smooth back reflector Ag, Optimally Doped ZnO. Pareto-front of \textit{OptIA-II} optimization is indicated in red.}
        \label{fig:select_rAg_oZnO_OPT}
	\end{figure}

    Our OptIA-II algorithm fine-tuned the Pareto-front of NSGA-II in this stage.
   	Figs.~\ref{fig:select_sAl_oZnO_OPT}, \ref{fig:select_sAg_oZnO_OPT}, and~\ref{fig:select_rAg_oZnO_OPT} show these fine-tuned Pareto-fronts.
   	Here, OptIA-II Pareto-fronts are shown in red line and feasible candidate solutions in coloured ``+'' marked points. 
   	NSGA-II Pareto-front solutions are shown in black ``+'' marked points.

   	OptIA-II improved the Pareto-front line of all three select NSGA-II obtained solutions.
   	Table~\ref{tab:simulation_select} shows the parameter values $ \boldsymbol{\sigma} $ and objective values $Q_e$ and $d_{\mu c-Si} $ of the fine-tuned solutions.

    The quantum efficiency, $Q_e$ of smooth BR Al plus optimally doped ZnO (sAl-oZnO) design increased from $Q_e = 0.5946$ to $Q_e = 0.6006$, i.e., its quantum efficiency, $Q_e$ improved by $1.0091\%$. 
    Also, the thickness $d_{\mu c-Si}$ of the  microcrystal silicon intrinsic layer decreased from $2210$ to $2208.43$. 
    Similarly, the quantum efficiency, $Q_e$ of smooth back reflector Ag plus optimally doped ZnO (sAg-oZnO) design increased from $Q_e = 0.5988$ to  $Q_e =0.6028$, a $0.6680\%$ increase in $Q_e$. 
    Its thickness $d_{\mu c-Si}$ decreased from $2210$ to $2208.24$. 
    For rough BR Ag plus optimally doped ZnO (rAg-oZnO) design, the quantum efficiency, $Q_e$ increased from $Q_e = 0.5990$ to $Q_e = 0.6031$. A $0.6845\%$ increased in quantum efficiency, $Q_e$. 
    Its thickness, $d_{\mu c-Si}$ decreased from 2210 to 2209.67.

    Notice that the energy production efficiency is directly proportional to increases in the quantum efficiency $Q_e$ of a single \textit{photovoltaic solar cell}. 
    Therefore, a $0.6845\%$ increased in  quantum efficiency $Q_e$ of a single \textit{photovoltaic solar cell} is a significant proportional improvement on the \textit{single solar panel} that has a few such photovoltaic solar cells.

    Our results in Tables~\ref{tab:simulation_sigma} and Fig.~\ref{fig:simulation_all_results} provide a detailed characterization of solar cell designs mentioned in Table~\ref{tab:simulation_detail}.
    This is significant for the characterization of solar cells real-world applications.

    For example, the applications such as household appliances and toys where a low-cost solar panel is required with relatively good quantum efficiency, we may use the least cost-intensive designs that have relatively good quantum efficiency. 
    In our obtained solar cell characterization, the second most-effective design cluster shown in Fig.~\ref{fig:simulation_all_results} pertaining to ``\textit{normally doped} SnO$_2$ and Al as a BR design may be used. 
    This design is highly cost-effective since normal doping is a cost-effective method than the optimally doped method, and TCO material SnO$_2$ is cheaper martial than ZnO. 
    Moreover, Al as a BR is cheaper than Ag.

    The design ``\textit{optimally} doped ZnO with Al as a (smooth) back reflector'' belonging to the most efficient cluster (see Fig.~\ref{fig:simulation_all_results} and its improved quantum efficiency in Table~\ref{tab:simulation_select}) is still a cost-effective design than the most effective design that is ``\textit{optimally} doped ZnO with Ag as a (rough) BR.'' 
    This design may be used in application relatively higher sophistication than the toys and household appliances. 
    Some sophisticated real-world applications such as satellites where highly efficient solar panels are required can use ``\textit{optimally} doped ZnO with Ag as a (rough) BR.'' 
        
	\begin{table}
		\small
		\centering
		\renewcommand{\arraystretch}{1.2}
		\setlength{\tabcolsep}{2.1pt}
		\caption{For each setting the points with the highest efficiency for the \emph{NSGA-II} and \emph{OptIA-II} optimizations are reported. Algorithm ``A'' indicate NSGA-II and algorithm ``B'' indicate NSGA-II + OptIA-II.}
		\label{tab:simulation_select}
		\begin{tabular}{llcccccccccccc}
			\toprule
			Simulation & Algo & $\sigma_2$ & $\sigma_3$ & $\sigma_4$ & $\sigma_5$ & $\sigma_6$ & $\sigma_7$ & $\sigma_8$ & $\sigma_9$ & $\sigma_{10}$ & $\sigma_{11}$ & $d_{\mu c-Si}$ & $Q_e$\\
			\midrule
			\multirow{2}*{sAl-oZnO} & {A} & 156.25 & 1.43 & 59.48 & 0.27 & 1.21 & 2.69 & 228.92 & 0.83 & 3.28 & 6.75 & 2210.00 & 0.5946\\
            & {B} & 179.86 & 3.13 & 59.80 & 3.94 & 3.13 & 3.85 & 131.39 & 3.04 & 3.84 & 27.28 & 2208.43 & 0.6006\\[1em]          
			\multirow{2}*{sAg-oZnO} & {A} & 180.00 & 1.72 & 60.00 & 0.66 & 1.48 & 2.46 & 39.03 & 1.25 & 0.75 & 11.86 & 2210.00 & 0.5988\\
			& {B} & 178.22 & 3.77 & 59.84 & 3.78 & 3.47 & 3.62 & 286.74 & 3.42 & 3.78 & 27.50 & 2208.24 & 0.6028\\[1em]
			\multirow{2}*{rAg-oZnO} & {A} & 180.00 & 1.38 & 56.95 & 0.38 & 0.88 & 3.19 & 211.80 & 1.26 & 2.96 & 6.44 & 2210.00 & 0.5990\\
			& {B} & 179.48 & 3.80 & 58.98 & 2.25 & 3.98 & 3.62 & 158.49 & 3.36 & 3.80 & 29.14 & 2209.67 & \textbf{0.6031}\\
			\bottomrule
		\end{tabular}
	\end{table}

    \clearpage
	\section{Conclusions
		\label{sec:con}}
	A comprehensive framework was designed in this research for solar cell optimization. 
	This framework studies 15 different solar cell structure designs.
	We performed simulation and optimization of cell structure interface roughness parameters to improve cells light-harvesting capacity. 
	That is, to improve cells quantum efficiency computed using a full Maxwell simulation model.

    We treated solar cell design optimization problem as a multi-objective optimization (MOO) problem and optimized cells quantum efficiency against cells microcrystalline silicon intrinsic layer thickness (cells fabrication cost). 
    Our Pareto-optimality results obtained by applying non-dominated sorting algorithm-II (NSGA-II) produced a \textit{full characterization} of the cell designs. 
    From the analysis of the NSGA-II produced Pareto-fronts, we found that transparent conductive oxide (TCO) layer doping has a strong correlation with cells quantum efficiency. 
    This means, in our experiments, a high concentration doping strategy (optimal doping with resistance less than \SI{1}{\meter\ohm}$\times$\SI{}{\centi\meter}) produced the most efficient cell design.

    We also found that the cell structure design with a \textit{rough back reflector} compared to a \textit{smooth back reflector} provided a higher light-harvesting capacity. 
    This was evident from the Pareto-front of cell designs for both zinc oxide (ZnO) and tin oxide (SnO$_2$) doping. 
    Here, the rough back reflector provided a better trad-off between quantum efficiency and micro-crystal silicon layer thickness. 
    
    The use of silver (Ag) as a back reflector material was clearly better than the aluminium (Al). 
    However, high concentration doping of ZnO with Al as back reflector material offered a slightly improved quantum efficiency ($0.6006$) with respect to the baseline quantum efficiency ($0.6$). 
    We found that ZnO as a TCO layer is more efficient than SnO$_2$.

    In our two-stage MOO, the Pareto-fronts of three select best cell designs of NSGA-II based MOO stage were fine-tuned by our designed multi-objective optimization-immunological algorithm (OptIA-II). 
    We observed that OptIA-II algorithm improved both costs associated with solar cell design. 
    It maximized quantum efficiency and minimized micro-crystal silicon intrinsic layer thickness of all three select designs. 
    The best stable solar cell design found was \textit{rough back reflector Ag plus optimally doped ZnO} that produced an $\approx 0.7$\% improved quantum efficiency (i.e., $Q_e = 0.6031$) with respect to the baseline quantum efficiency.
    
    
	\bibliographystyle{IEEEtran} %
	\bibliography{SolarCells}
	


\end{document}